\documentclass{svjour3}                     

\smartqed  

\usepackage{graphicx}

\begin{document}


\title{Instruments of RT-2 Experiment onboard CORONAS-PHOTON
and their test and evaluation IV: Background Simulations using GEANT-4 Toolkit
\thanks{This work was made possible in part from a grant from Indian Space
Research Organization (ISRO). The whole-hearted support from G. Madhavan Nair,
Ex-Chairman, ISRO, who initiated the RT-2 project, is gratefully acknowledged.}
}

\titlerunning{RT-2 payloads aboard CORONAS-PHOTON IV: Geant-4 simulation}

\author{Ritabrata Sarkar \and Samir Mandal \and Dipak Debnath \and Tilak B. Kotoch 
\and Anuj Nandi \and A. R. Rao  \and Sandip K. Chakrabarti}

\authorrunning{Ritabrata Sarkar et. al. } 

\institute{Ritabrata Sarkar, Samir Mandal, Dipak Debnath, Tilak B. Kotoch,
Anuj Nandi$^+$\at
Indian Centre for Space Physics, 43 Chalantika, Garia Station Rd.,
Kolkata 700084\\
\email{ritabrata@csp.res.in; samir@csp.res.in; dipak@csp.res.in; tilak@csp.res.in; 
anuj@csp.res.in}\\
($+$: Posted at ICSP by Space Science Division, ISRO Head Quarters)
\and
A. R. Rao \at
Tata Institute of Fundamental Research, Homi Bhabha Road, Colaba, 400005\\
\email{arrao@tifr.res.in}
\and
Sandip K. Chakrabarti \at
S.N. Bose National Centre for Basic Sciences, JD Block, Salt Lake, Kolkata 700097\\
(Also at Indian Centre for Space Physics, 43 Chalantika, Garia Station Rd.,
Kolkata 700084)\\
\email{chakraba@bose.res.in}           
}

\date{Received: date / Accepted: date}

\maketitle


\begin{abstract}

Hard X-ray detectors in space are prone to background signals due to the
ubiquitous cosmic rays and cosmic diffuse background radiation that continuously
bombards the satellites which carry the detectors. In general, the background
intensity depends on the space environment as well as the material surrounding
the detectors. Understanding the behavior of the background noise in the 
detector is very important to extract the precise source information from 
the detector data. In this paper, we carry out Monte Carlo
simulations using the GEANT-4 toolkit to estimate the prompt background noise 
measured with the detectors of the RT-2 Experiment onboard the CORONAS-PHOTON 
satellite.



\end{abstract}

\section{Introduction}

Observational astronomy in the X-ray and $\gamma$-ray bands of the
electromagnetic spectrum is very crucial to explore high-energy physical
phenomena in the Universe. X-ray or $\gamma$-ray observations from the ground-based
instruments are not possible due to the atmospheric attenuation. In the last
four decades, huge efforts have been made towards the development of
space-borne X-ray and $\gamma$-ray telescopes. At the same time, these efforts
are also limited by the hostile space environment, particularly in hard
X-rays and $\gamma$-rays. High energy charged particles coming from outer
space and from the Solar wind become trapped in the Earth's surrounding magnetic
field creating radiation belts around the Earth, known as the Van Allen radiation
belt. Though the satellites carrying the X-ray and $\gamma$-ray detectors are
usually placed below the inner radiation belt (altitude varying from a few
$100 ~km$ to $10,000 ~km$), still there are some localized energetic charged
particle regions which may result in severe damages to the instruments if they
are activated while passing through those regions such as the Polar Regions and
the South Atlantic Anomaly (SAA) region. Apart from these trapped charged particle
regions, there are the cosmic diffuse radiation and the
cosmic rays, mainly protons and alpha particles.
These cosmic-ray particles, depending on the geomagnetic strength at the altitude and 
position, enter into the Earth's atmosphere and interact with the atoms and molecules 
resulting in various secondary particles. These primary and secondary cosmic radiations 
and charged particles bombard the detector and satellite materials and may
produce secondary or higher order particles in prompt interactions (through bremsstrahlung, 
pair creation etc.) or/and may initiate the detector activation. These radiations and
particles will increase the detector noise which may mask the original source signal. The 
high-energy charge particles can pass through detector and space craft and may deposit a
line of charge in the detector volume. This may then be detected in the same way as energy 
deposit produced by the `real' X-ray radiations.
It is thus very essential to make an accurate estimate of the onboard background
noise before designing any space experiment. The background noise in an X-ray
instrument is mainly due to the interactions of the cosmic-ray protons, the
albedo protons and neutrons due to Earth's atmosphere, Cosmic Diffused Gamma-Ray
Background (CDGRB), secondary gamma rays formed in the detector material, its
frame structure and with the satellite (Dean, Lei \& Knight 1991). In addition, the long-term 
activation of the detector materials by these radiations or particles or by the particles
in the trapped particle region are also responsible for the background noise.
The background noise varies over a wide range of energy and actually depends
on the detection capability of the specific detector.  These background noise
compete with the signals due to the interactions of source photons with the
detectors. It is therefore important to understand the interactions of these
background components with the detector material and to remove them while
extracting the source signal.

The RT-2 experiment aboard the CORONAS-PHOTON satellite (Kotov et al. 2008,
Nandi et al. 2009) consists of 4 payloads: three X-ray detectors (RT-2/S,
RT-2/G \& RT-2/CZT) and one processing electronic device (RT-2/E). Detailed
description of all the payloads and their functionality are given in Debnath
et al. (2010), Kotoch et al. (2010), Nandi et al. (2010) and Sreekumar et al. (2010).
The Phoswich detectors (RT-2/S \& RT-2/G) are made of NaI (Tl) and CsI (Na)
scintillating crystals. Both the Phoswich detectors are sensitive to detect
high energy X-rays in the energy range of $15 ~keV$ to $\sim 1000 ~keV$. The
RT-2/CZT detector is a solid-state imaging device, which consists of CZT and
CMOS detectors. Both the detectors are sensitive in the energy range of
$20 ~keV$ to $150 ~keV$.

In the present work, we concentrate on predicting/comparing the background in
these detectors during the passage of the space craft through the low-background
equatorial region (away from the SAA and polar regions), to estimate the sensitivity of 
the detector. Hence, we make a detailed simulation of the interaction of primary and 
secondary protons, cosmic diffused gamma rays, secondary gamma rays and secondary
neutrons with the detector volume as well as the whole structure of the
satellite carrying the detectors.

In this paper, we carry out Monte Carlo (MC) simulations using the GEANT-4
toolkit and highlight the effects of shielding material in calculating the
background noise due to cosmic-ray photons on the detectors of the RT-2
Experiment. In the next section (\S 2), we describe the typical spectrum of
the CDGRB, primary cosmic-ray proton and secondary proton, gamma-ray and
neutron spectrum in the low Earth orbit of interest and of a standard GRB
source. We also discuss the physical processes that are involved while the incident 
particles and radiations interact with the detector material. In \S 3 \& \S 4,
we present the simulation results of the RT-2/S (RT-2/G) and RT-2/CZT payloads.
In \S 5 we compare the predicted result with an observed data set. Finally, we
conclude in section \S 6.

\section{The Simulation Attributes}

\subsection{Detector characteristics \& specifications}

The RT-2 Experiment aboard the CORONAS-PHOTON satellite is a unique experiment 
as it consists of different types of detectors, namely a scintillator detector 
(NaI, CsI crystal), a solid-state detector (CZT) and a photo-diode (CMOS) 
detector. The specifications and materials used for all these detectors are 
given in Table 1.

\begin{table}[htp]
\noindent{Table 1: Detector specifications of RT-2/S (RT-2/G) and RT-2/CZT payloads.}
\vskip 0.5cm
\scriptsize
\centering
\begin{tabular}{lcc}
\hline
{\bf Payload}        & {\bf RT-2/S (RT-2/G)} & {\bf RT-2/CZT} \\
\hline
Detector type        & NaI + CsI             & CZT \& CMOS \\
Material composition & NaI (Tl activated)    & Cd$_{0.9}$Zn$_{0.1}$Te \\
		             & CsI (Na activated)    & CMOS (Amorphous Si photo-diode \\
                     &                       & Gd$_2$O$_2$S:Tb) \\
Thickness (mm)       & 3 + 25                & 5 \& 3 \\
Size (mm)            &  116 dia              &  40$\times$40 \& 24.5$\times$24.5 \\ 
Filter (mm)          & Al (0.5)              &  graphite (1.0) (protective cover) \\
Effective area (cm$^2$) & 105.6              & 48 (3 modules) \& 5.7 \\
FOV                  & $4^\circ$ $\times$ $4^\circ$ ($6^\circ$ $\times$ $6^\circ$) & $6'$ $-$ $6^\circ$ \\
Readout              & PMT                   & pixels \\
\hline
\end{tabular}
\end{table}

For the simulation of the effects of various background components in the 
detectors we consider an approximate mass distribution of the whole 
CORONAS-PHOTON satellite hosting various detectors on it. Figure 1 depicts
the simplified shape of the whole satellite including the detectors on it.
This shape was used in the simulation. The major contribution
of its mass is from the satellite shell structure which is made up of Aluminum (Al) and
the electronics modules inside it, which consist of Aluminum (Al), Silicon (Si) and 
Copper (Cu) as the major elements. We distribute these materials throughout the satellite cavity
for simplification. Also for the detector components other than
RT-2/S, RT-2/G and RT-2/CZT we consider simplified structures consisting of
the approximate weights of the major components of concerned detectors. The
overall height of the satellite construction under our consideration is
$\sim 324 ~cm$ and radius is $\sim 85 ~cm$. For the detectors of our concern
i.e., RT-2/S, RT-2/G and RT-2/CZT, we use more detailed geometry described in 
the following sections.

\begin {figure}[htp]
\centering{
\includegraphics[width=10.0cm]{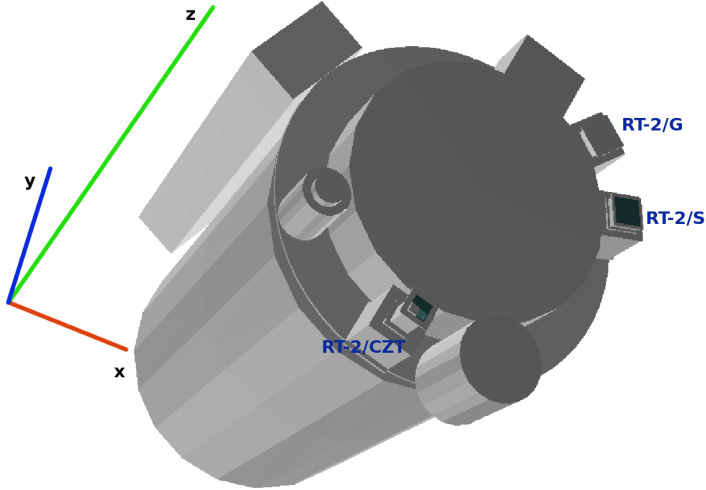}
\caption{A simplified 3D view of the approximate mass distribution of the 
satellite containing the detectors used for the simulation.}}
\end {figure}

\subsection{Primary Particle Generation}
 
We simulate five major components which cause background noise in X-ray or 
$\gamma$-ray detectors. They are: the CDGRB photons, secondary gamma-ray
photons due to Earth's atmosphere, primary Cosmic-Ray (CR) protons, secondary protons 
due to interaction of CR in Earth's atmosphere and the albedo neutrons from the 
Earth's atmosphere.

To simulate the CDGRB photons, we generate the incident photons from the
surface of a hemisphere (at +ve z-axis) of radius $120.0 ~cm$. The center
of the whole geometry is at the center of the satellite common mounting plate on which 
the RT-2 detectors and some other detectors are mounted (see Figure 1). The position
of the incident photon generating hemisphere is such that it covers the whole 
region above the Earth's horizon. The randomness of the incident photons are ensured by 
placing their origin randomly on the hemisphere and the directions of the photon momenta 
are also chosen to be random within the solid angle subtended by the dimension of the 
satellite radius at the vertex of each incident photon.
We are interested in the response of the detector in the incident energy range
of $10 ~keV-100 ~MeV$. The CDGRB spectrum can be represented by the equation
(Gruber et al. 1999),

\begin{equation}
\frac{dN}{dE} = \left\{
\begin{array}{l l}
7.877E^{-1.29}\exp^{-E/41.13} & \quad \mbox{if $E\leq 60.0~keV$}\\
4.32 \times 10^{-4}\left(\frac{E}{60}\right)^{-6.5} + 8.4 \times 10^{-3}\left(\frac{E}{60}\right)^{-2.58} \\
+ 4.8 \times 10^{-4}\left(\frac{E}{60}\right)^{-2.05} & \quad \mbox{if $E\geq 60.0~keV$},
\end{array} \right. 
\end{equation}
where, $E$ is incident photon energy and $\frac{dN}{dE}$ is in the unit of 
$counts/cm^2/s/sr/keV$. We divide the entire energy range into $500$ bins
equal in logarithmic scale and simulate $100,000$ photons in each bin to retain
a good statistics in the simulation result. The incident photon spectrum on the
detectors due to CDGRB is shown in Figure 2a.

For the simulation of the secondary albedo gamma-ray photons, we consider photons
randomly originated from a hemispherical surface of radius $300 ~cm$ with the
center coinciding with the center of the satellite common mounting plate.
In this case, we consider the position of the generating hemisphere at the
opposite that of the primary CDGRB (i.e., at -ve z-axis). The direction of the 
secondary photons has been achieved in the same way as that of the primary CDGRB.
In this case also we consider the energy range of $10 ~keV-100 ~MeV$. 
Atmospheric gamma-ray line emission, such as the $511 ~keV$ emission from positron 
annihilation, has not been considered, while simulating for secondary albedo 
gamma-ray photons. The energy spectrum is represented by (Ajello et al. 2008;
Mizuno et al. 2004),
\begin{equation}
\frac{dN}{dE} = \left\{
\begin{array}{l l}
\frac{1.87 \times 10^{-2}}{\left(\frac{E}{33.7}\right)^{-5.0} + \left(\frac{E}{33.7}\right)^{1.72}} & \quad \mbox{if $E\leq 200.0~keV$}\\
1.01 \times 10^{-4} \left(\frac{E}{MeV}\right)^{-1.34} & \quad \mbox{if $200.0 ~keV \leq E\leq 20.0~MeV$}\\
7.29 \times 10^{-4} \left(\frac{E}{MeV}\right)^{-2.0} & \quad \mbox{if $E\geq 20.0~MeV$}\\
\end{array} \right.
\end{equation}
in units of $counts/cm^2/s/sr/keV$. We divide the entire energy range into 
$500$ bins equal in logarithmic scale and in each bin, we inject $100,000$ photons. 
The incident photon spectrum on the detectors due to secondary gamma-ray photons 
is shown in Figure 2b.

We also simulate the detector response due to the CR and secondary protons. 
We consider the input differential spectra for the downward and upward going protons.
While the upward going protons are mostly secondaries from the Earth's atmosphere,
the downward going component also contains the primary CR proton above the cutoff.
We consider these spectra near the equatorial region.
 
For the downward going proton component we consider random protons from a spherical
section of radius $300 ~cm$ around the satellite with its center coinciding 
with the center of the common mounting plate and covering the open region 
above the Earth's horizon. The randomization in the direction of 
the particles is the same as described for the CDGRB or secondary photons. We carry 
out the simulation for the proton energy range of $100 ~MeV-20 ~GeV$. To 
produce the energy distribution we consider the spectral data given by Alcaraz et al. 
(2000) for the low geomagnetic latitude ($0 < \Theta_M < 0.2$). We fit this data 
using the functional form described by Mizuno et al. (2004) presented by the 
equation,

\begin{equation}
\begin{array}{l l}
\frac{dN}{dE} = & 1.23 \times 10^{-8} \left(\frac{E}{GeV}\right)^{-a} \exp - \left(\frac{E}{E1_{cut}}\right)^{-a+1} \\
& + 16.9 \times 10^{-7} \left(\frac{E+Ze\phi}{GeV}\right)^b \times \frac{(E+Mc^2)^2-(Mc^2)^2}{(E+Mc^2+Ze\phi)^2-(Mc^2)^2} \times \frac{1}{1+\left(\frac{E}{E2_{cut}}\right)^{-12.0}}
\end{array}
\end{equation}
in units of $counts/cm^2/s/sr/keV$, where $e$ is the magnitude of the electron
charge, $Z$ is the atomic number of the particle, $a=0.155$, 
$E1_{cut}=5.1 \times 10^5 ~keV$, Solar modulation $\phi = 6.5 \times 10^5 ~kV$ 
(a value near Solar activity minimum), $Mc^2$ is the proton 
mass and $E2_{cut} = 12.25 \times 10^6 ~keV$.
In this case, we divide the whole incident energy range in $100$ bins equal
in logarithmic scale and inject $100,000$ photons in each bin. Figure 2c presents 
the incident energy spectrum for the Primary CR protons.

The contribution of the upward going proton spectra from the Earth's 
atmosphere is considered as follows. For the positional and directional aspects 
of the generation of these protons we used the same methods as the albedo 
photons. Here in this simulation we consider the energy range of $100~ MeV - 
6 ~GeV$. The spectral form is given as (Alcaraz et al. 2000; Mizuno et al. 2004),

\begin{equation}
\frac{dN}{dE} = 1.23 \times 10^{-8} \left(\frac{E}{GeV}\right)^{-a} \exp - \left(\frac{E}{E_{cut}}\right)^{-a+1}
\end{equation}
in units of $counts/cm^2/s/sr/keV$, where, $a=0.155$ and $E_{cut}=5.1 \times 10^5 ~keV$.
The incident secondary proton spectrum is depicted in Figure 2d.

For the simulation of the secondary neutrons due to the interaction of the 
CR in the Earth's atmosphere, we consider the generation of the neutrons 
from a hemispherical surface of radius $300 ~cm$ in the same manner as the
secondary photon simulation. The incident particle direction is also 
randomized in the same way. The energy range of the neutron simulation is 
$10 ~keV - 1 ~GeV$. The spectral form of the neutron energy distribution
is given by (Armstrong et al. 1973),

\begin{equation}
\frac{dN}{dE} = \left\{
\begin{array}{l l}
9.98 \times 10^{-8} \left(\frac{E}{GeV}\right)^{-0.5} & \quad \mbox{if $10~keV \leq E \leq 1~MeV$}\\
3.16 \times 10^{-9} \left(\frac{E}{GeV}\right)^{-1.0} & \quad \mbox{if $1~MeV \leq E \leq 100~MeV$}\\
3.16 \times 10^{-10} \left(\frac{E}{GeV}\right)^{-2.0} & \quad \mbox{if $100~MeV \leq E \leq 100~GeV$}\\
\end{array} \right.
\end{equation}
in units of $counts/cm^2/s/sr/keV$. We divide the whole energy range in $500$ energy bins 
equal in log scale and inject $100,000$ photons in each bin. Figure 2e shows the incident
spectrum for the secondary neutron generation.

\begin {figure}[htp]
\vbox{
\centering{
\includegraphics[width=0.49\textwidth]{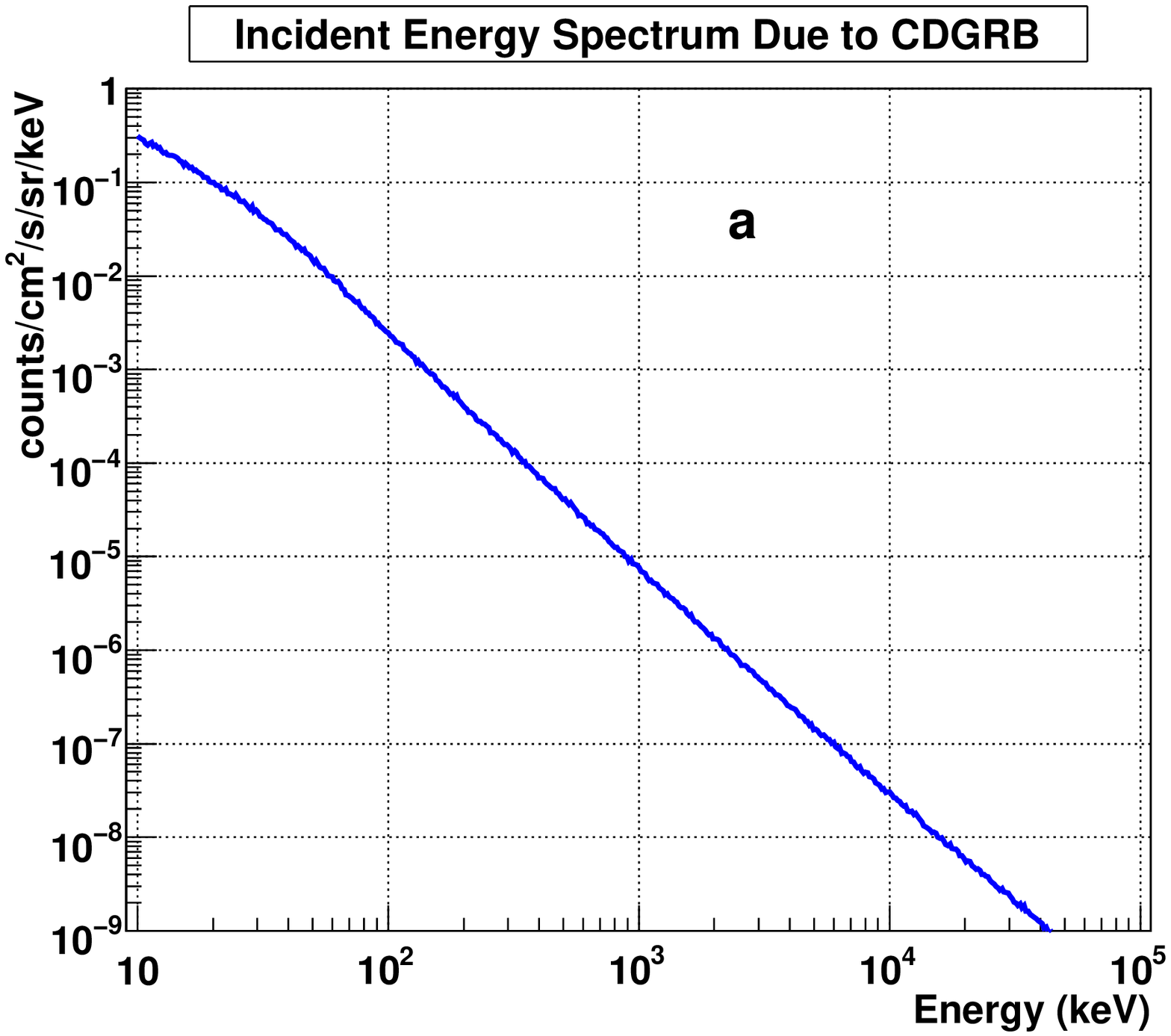}
\includegraphics[width=0.49\textwidth]{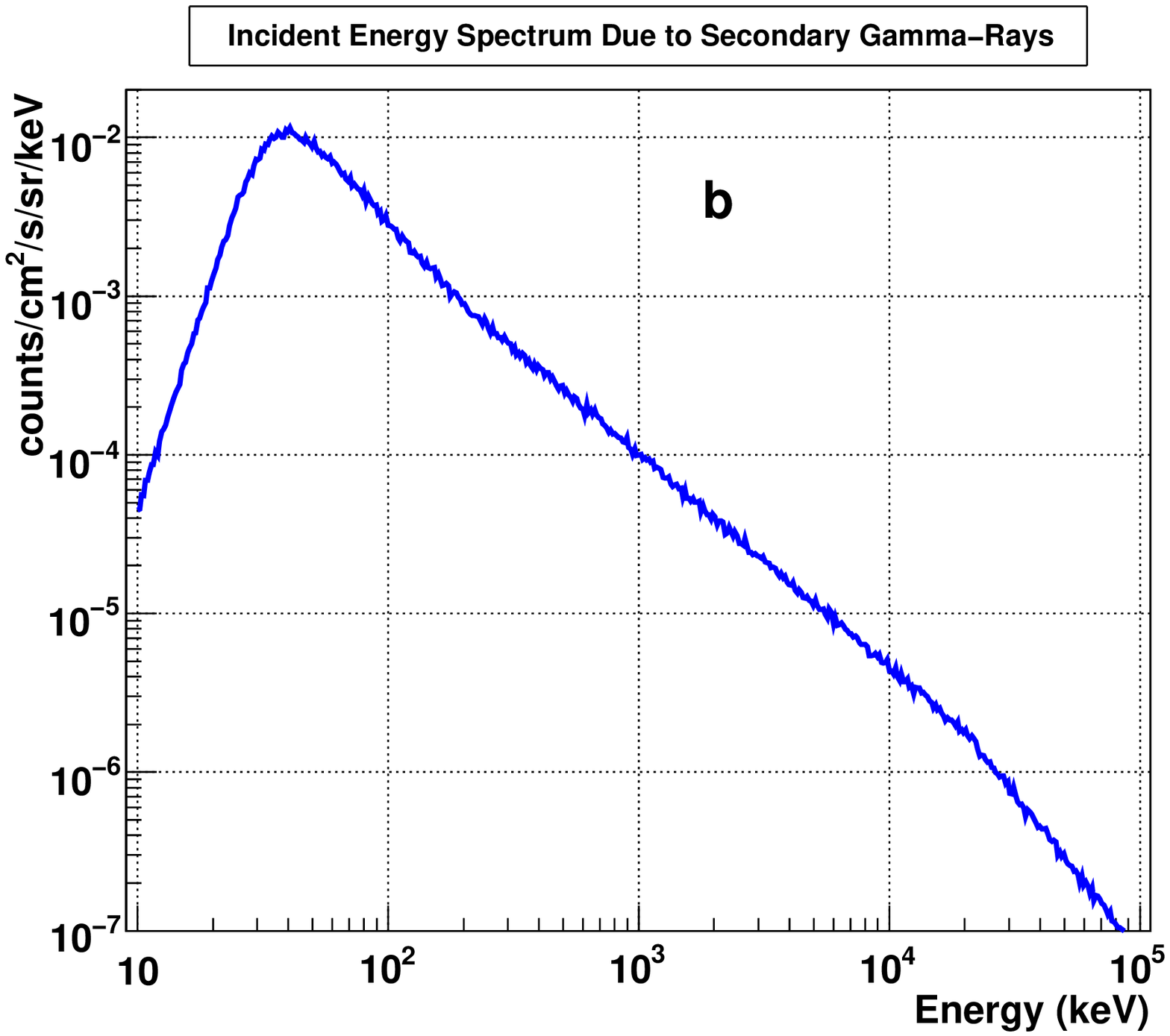}
\includegraphics[width=0.49\textwidth]{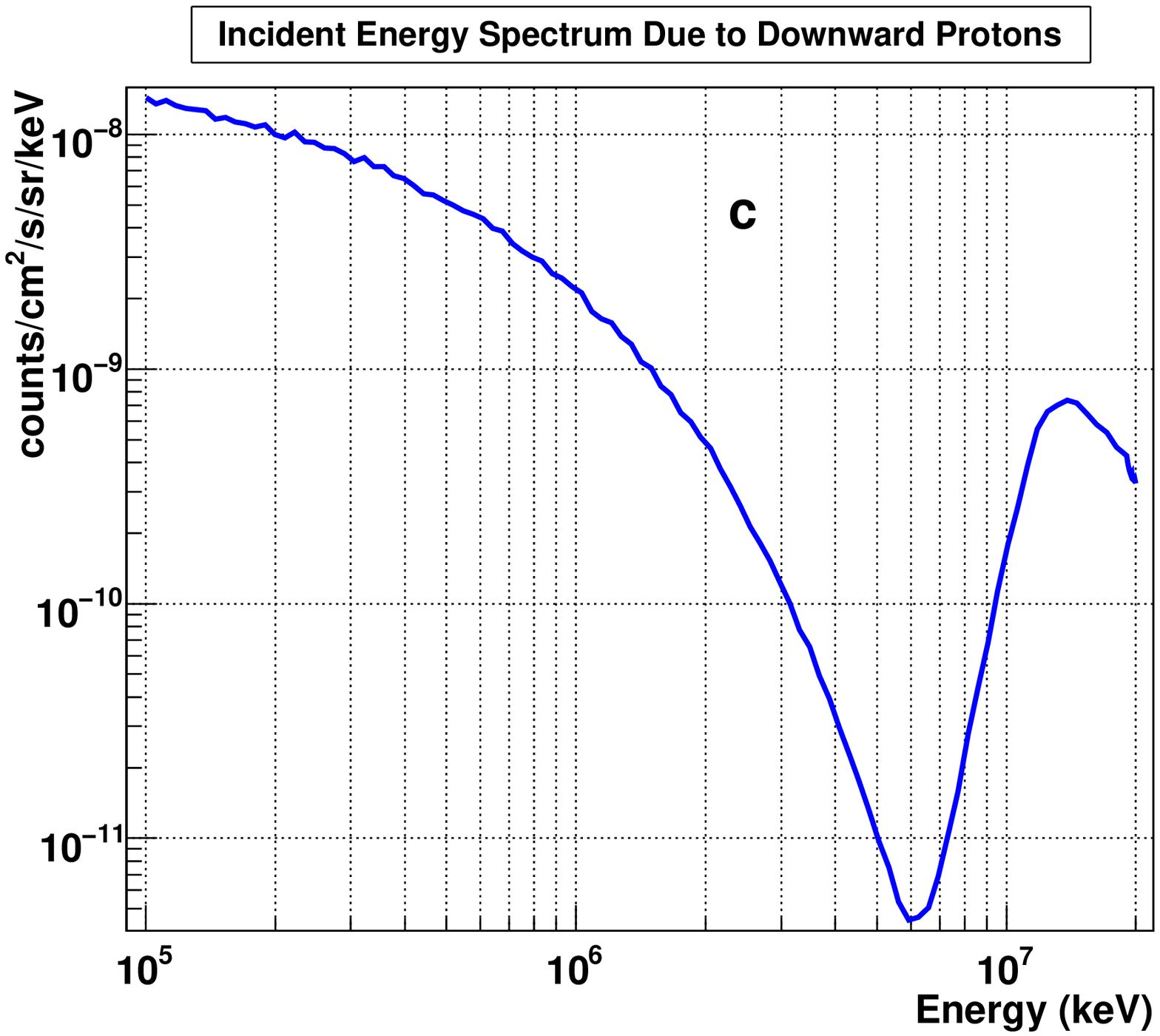}
\includegraphics[width=0.49\textwidth]{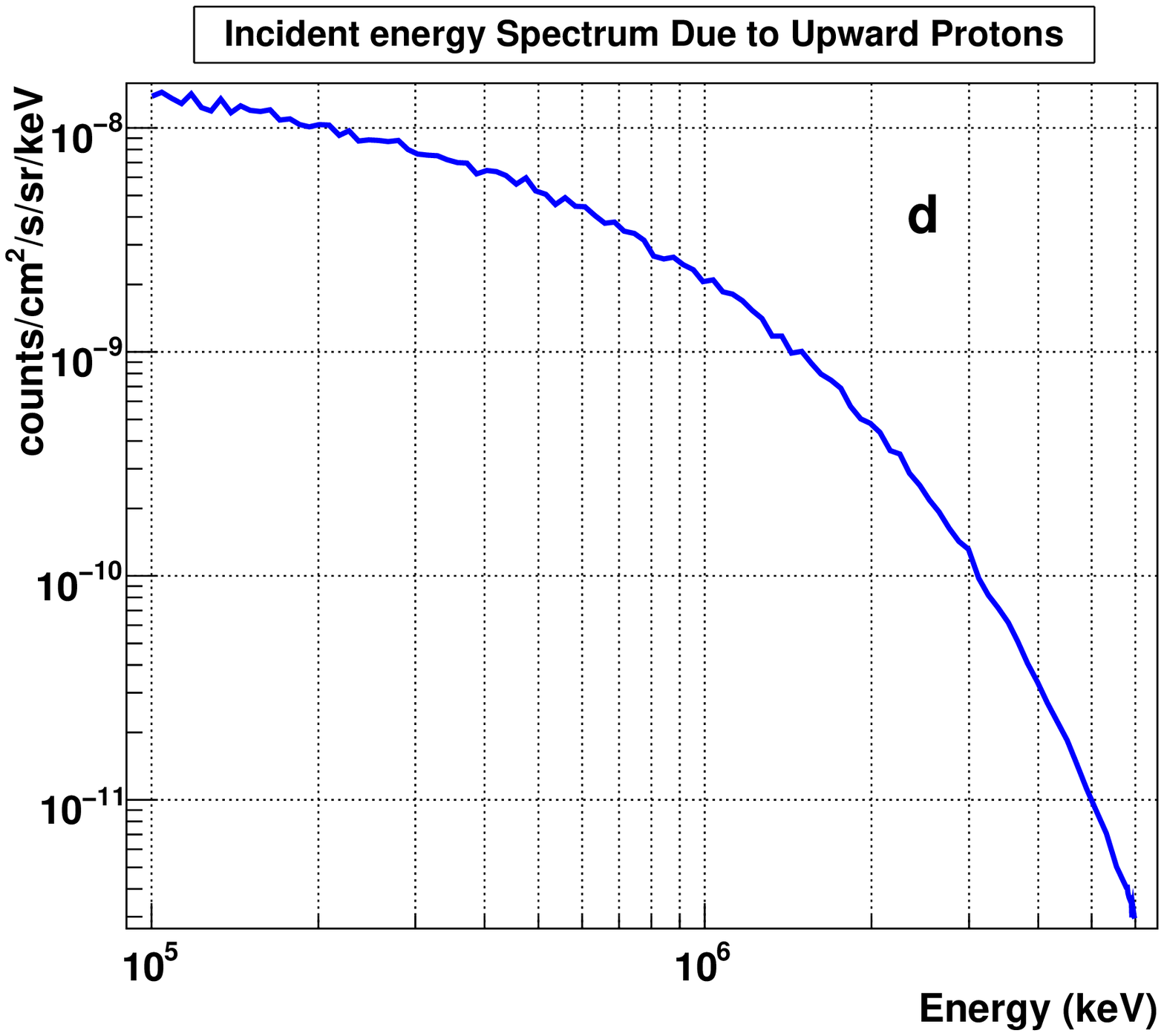}
\includegraphics[width=0.49\textwidth]{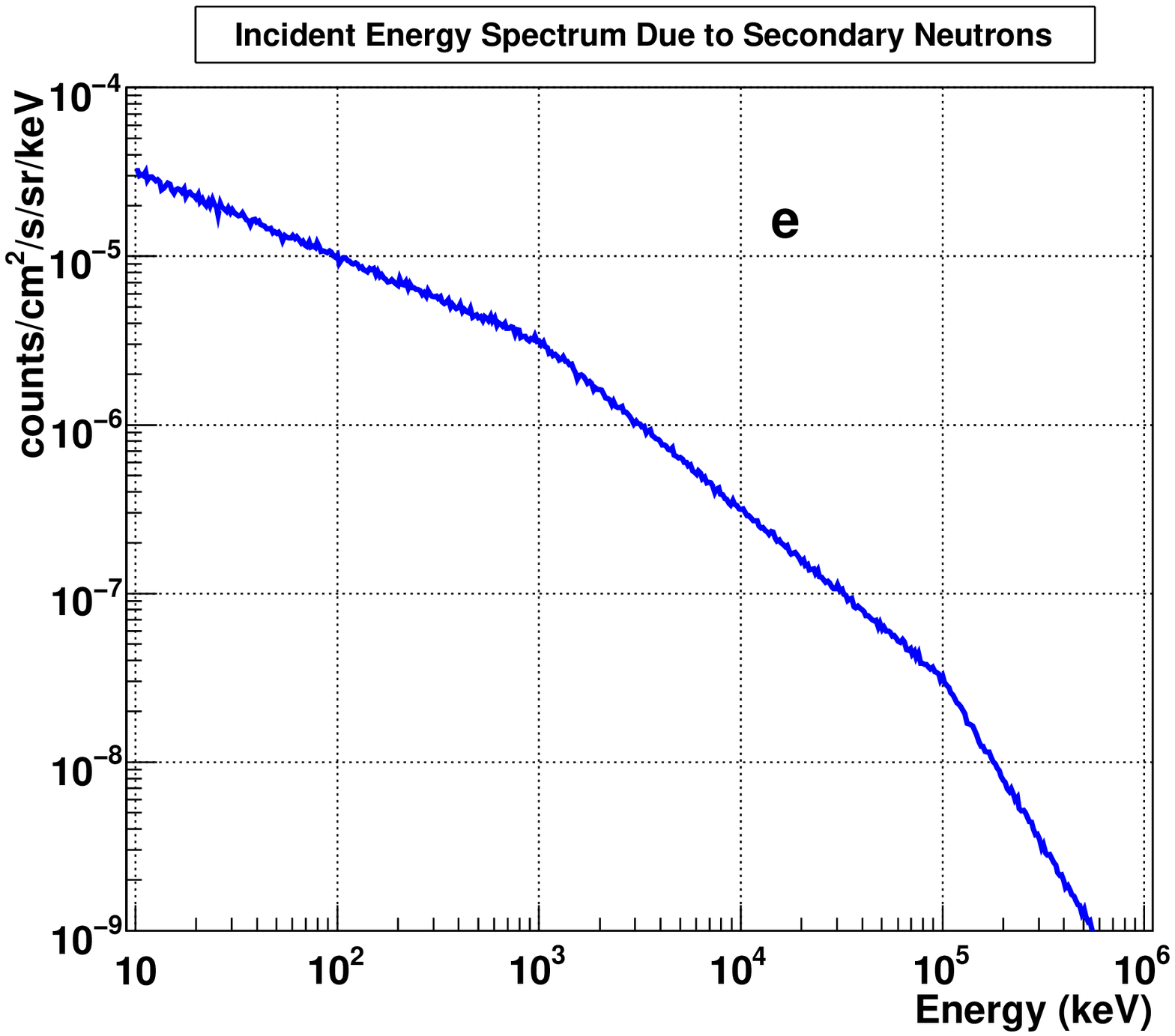}
\caption{The incident energy spectrum due to the (a) CDGRB photons, 
(b) secondary gamma-ray photons, (c) downward going protons, (d) upward going
protons and (e) albedo neutrons.}}}
\end {figure}

The spectrum of energy deposition on the detectors is calculated by normalizing
the deposited spectrum in the following way. We have $N$ incident energy bins
equal in logarithmic scale, where $E_i$ is the width of the $i$th bin. In each
bin we are simulating $I_s$ (here $100,000$) number of particles from a 
surface of area $A_I$ and over the solid angle $\Omega_I$ 
subtended by the detector to the vertex of the incident photons. Now the total
number of photons in each incident energy bin is 
$I_i = \int_{E_i}\int_{A_I}\int_{\Omega_I} \frac{dN}{dE} dE$. Then we calculate the 
{\it normalization constant} $C_i = \frac{I_i}{I_s}$. To calculate the 
normalized deposition spectrum we divide the whole deposition energy range in 
$M$ bins. We have $D_{ij}$ number of photons in the $j$th bin of the 
deposition spectrum due to $I_s$ photons in the $i$th incident bin. Then
we calculate the normalized photon counts in the $j$th bin of the deposition
spectrum for the all incident photons as 
\begin{equation}
D_j = \sum_{i=1}^N \frac{D_{ij} \times C_i}{A_D \times E_j}
\end{equation}
where, $A_D$ is the area of the crystal (detector) and $E_j$ is the width
of the $j$th energy range.

In the present work, we are only dealing with the sources for the 
prompt background noise. So we simulate here some of the main sources of 
the prompt background noise sources. Apart from the prompt background, 
significant noise will be present due to the detector material 
activation as the satellite is in a polar orbit. However, there are 
considerable uncertainties in estimating the contribution due to long term activation 
to the total background. This is because the relevant package, namely, {\it Cosima}
based on Geant4 is still in the developmental stage and cannot be 
trusted for predicting backgrounds due to activation (Zoglauer et al. 2008; Zoglauer, 2009). 
In the present paper, we have deferred the inclusion of the effects 
of activation on the detector. This can be dealt with in a future work.

To simulate the incident spectrum of a Gamma-Ray Burst (GRB), we consider 
the Band spectral (Band et al. 1993) form of a bright GRB (GRB 880725), 
which follows the equation, 
\begin{equation}
\frac{dN}{dE} = \left\{
\begin{array}{l l}
A\left(\frac{E}{100keV}\right)^{\alpha}\exp\left({-\frac{E}{E_0}}\right) & \quad \mbox{if $E\leq 
(\alpha-\beta)E_0$}\\
A\left(\frac{(\alpha-\beta)E_0}{100keV}\right)^{(\alpha-\beta)}\exp\left({\beta-\alpha}\right)
\left(\frac{E}{100keV}\right)^{\beta} & \quad \mbox{if $E\geq (\alpha-\beta)E_0$},
\end{array} \right.
\end{equation}
in units of $counts/cm^2/s/keV$, where, $A=7.056$ is a constant, $\alpha=-0.32$ is the low
energy slope, $\beta=-5.0$ is the high energy slope and $E_0=67.7~keV$ is the
break energy (Strohmayer et al. 1998). The incident photons are generated from
a plane of area equal to that of the collimator of the detectors and is placed
$60 ~cm$ above the detector base plate. We consider parallel photons falling straight into
the detector through the collimator. We also consider the source location
(GRB) at an angle of $\sim 50^\circ$ to the on-axis of the payload and its
effect on resulting photon distribution after interacting with detector materials. 
The incident energy spectrum (Eqn. 7) of the GRB is shown in Figure 3.

\begin {figure}[htp]
\centering{
\includegraphics[width=0.5\textwidth]{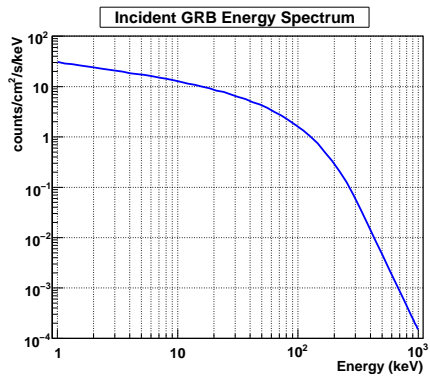}
\caption{The incident photon spectrum due to a typical Gamma-Ray Burst (GRB).}}
\end {figure}

\subsection{Physical Processes for Primaries as well as Secondaries}

The CORONAS-PHOTON satellite is placed at an altitude of $550 ~km$ in an orbit
which is inclined at  $82.5^\circ$, which results in the passage of the
satellite through high energetic charge particle regions of the SAA, North Cap 
(NC) and South Cap (SC) for $\sim40\%$ of its orbital time. In these regions,
satellite operations will be restricted for the protection of the detectors
from the high radiation dose. Apart from the high energetic charge particles
of SAA, NC and SC regions, the CDGRB, the high energy protons and other
albedo particles and photons will hit the satellite and detector 
materials to produce the background noise in the  detectors either directly
or by creating local spallation background.
The hard X-ray solar flares along with some other astrophysical sources
(e.g,. GRBs) would be detected by the RT-2 detectors in the energy range of 
$15-150 ~keV$, extendable up to $\sim1000 ~keV$. The underlying 
physical processes by which photons can interact in any medium depend on 
particle energy and the material properties of that medium.

For the simulation of the detectors under this circumstances, we use 
the Geant4 simulation toolkit version 9.1.p03 and the cross section data
version G4EMLOW5.1, G4ABLA3.0, G4NDL3.12. For the electromagnetic processes in
the simulation we consider the low energy electromagnetic physics list. For 
the photon interactions we are using the low energy photo-electric effect 
(activated with Auger electron production), low energy Compton effect, low 
energy Rayleigh Scattering and low energy Gamma ray conversion. In the above 
mentioned energy range, the incident photon produces electrons as the 
secondary particles. Electrons lose their  energy through the low energy 
ionization, multiple scatterings and low energy bremsstrahlung processes.
For positrons we consider the bremsstrahlung, annihilation, ionization 
and multiple scattering processes. We are using a production cut-off value 
of $1 ~\mu m$ for photon, electron and positron which is the same for all 
the materials. This implies a threshold energy in the Aluminum as $990 ~eV$, 
$1.1 ~keV$ and $1.1 ~keV$ respectively for photons, electrons and positrons. 
For the hadronic interactions we are considering the ``LHEP'' physics list
defined in the Geant4. In the Geant4 toolkit, above mentioned processes are
defined as separate modules in the ``process'' category and this module is
included in the ``Physics list''(Agostinelli et al. 2003). Further information
are available at http://geant4.web.cern.ch/geant4/.

\subsection{Shielding material}

To shield the detectors from X-rays and gamma-rays from off-axis sources and 
diffuse X-rays and other cosmic particles, materials with high atomic 
numbers such as Lead and Gold etc. are used. These high-Z elements, however, have a deep 
dip in their absorption coefficient just below their K-shell binding energies. Sometimes, 
materials with lower atomic number like Tin and Copper are used to absorb the
characteristic fluorescent X-rays. In the present case, however, the primary
objective of the experiment is to make very sensitive hard X-ray measurement
of solar flares (typically below $50 ~keV$) and in particular use the  Phoswich 
technique for background reduction. Since the CsI detector in the Phoswich
configuration  has a good sensitivity above $100 ~keV$, an alternate objective
is to use the detector as an open monitor above $100 ~keV$ for gamma ray bursts 
and other bright sources. For these dual objectives, we find that Tantalum has 
a very good absorption property with atomic number $73$, a K-shell binding 
energy of $67.4 ~keV$ and high density of $16.65 g cm^{-3}$.
For the shielding of the detector we have chosen the Tantalum shield 
in a way which minimizes mass as well as maximizes the dual objectives of 
solar X-ray flare spectroscopy and off-axis source observations.

\section{Simulation of RT-2/S and RT-2/G payload}

The background simulation of the RT-2/S (RT-2/G) detector is carried out with
the virtual detector constructed within the Geant4 toolkit environment with 
the construction parameters given in Table 2. These parameter values are the
same as in the onboard detector geometry, but to avoid the complexity in the
detector geometry construction in the simulation we simplified some of the
detector parts which are not expected to change the simulation results
significantly. These simplifications consist of omitting the ribs to hold the
collimator and the detector walls, simplification of the construction of the
Photo-Multiplier Tube (PMT), etc. A 3D view of the RT-2/S detector which is
considered in the simulation is given in Figure 4. This figure is the direct 
output from the simulation and agrees closely with the CAD design of the actual
detector construction. We simulate the detector response for the background 
noise due to five major background components and a typical GRB spectrum as
has been described earlier. In the following subsection, we will discuss
the response of the detector for these particular background radiation under 
our consideration and in the subsequent subsection we will give the results for
the GRB spectrum.

\begin {figure}[htp]
\centering{
\includegraphics[width=12.0cm]{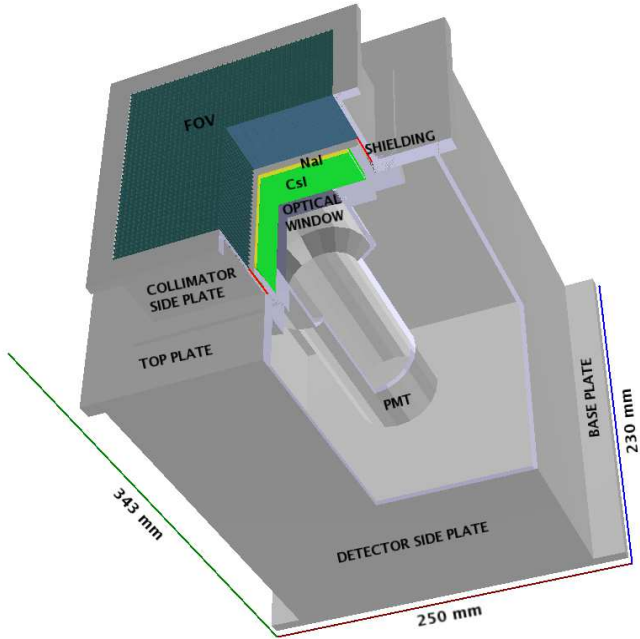}
\caption{A schematic 3D view of RT-2/S (RT-2/G) payload with different
components (e.g., crystals, collimator, PMT etc.). }}
\end {figure}

\begin{table}[htp]
\noindent{Table 2: Detector construction parameter specifications of RT-2/S
(RT-2/G) payload.}
\vskip 0.5cm
\scriptsize
\centering
\begin{tabular}{|l|c|}
\hline
{\bf Detector Parts} 	& {\bf Dimensions} \\
\hline
Bottom plate 			& $25.0\times23.0\times0.8~cm^3$ \\
\hline
Detector Housing 		& $20.2\times22.4\times26.7~cm^3$ (inner dimension)\\
                 		& wall thickness $0.3~cm$ \\
\hline
Top Plate 			& $23.0\times23.0\times1.0~cm^3$ \\
          			& hole radius $6.5~cm$ \\
\hline
Collimator 			& $13.25\times13.25\times5.8~cm^3$ (inner dimension)\\
           			& wall thickness $0.3~cm$ \\
\hline
FOV 				& $0.02~cm$ Tantalum plate thickness\\
    				& $0.4~cm$ gap between two Ta plates ($4^{\circ}$ $\times$ $4^{\circ}$ FOV for S) \\
    				& $0.6~cm$ gap between two Ta plates ($6^{\circ}$ $\times$ $6^{\circ}$ FOV for G) \\
\hline
NaI Crystal 			& radius $5.8~cm$, thickness $0.3~cm$ \\
\hline
CsI Crystal 			& radius $5.8~cm$, thickness $2.5~cm$ \\
            			& (gap between two crystals $0.01~cm$) \\
\hline
Optical Coupling 		& (Silicon Oxide) radius $3.8~cm$ thickness $1.25~cm$ \\
\hline
PMT 				& Upper part : radius $4.16~cm$, height $3.826~cm$ \\
    				& Middle part (conic section) : height $1.024~cm$ \\ 
    				& Lower part : radius $2.94~cm$, height $12.2~cm$ \\
    				& thickness of the whole PMT $0.36~cm$ (Aluminum) \\
\hline
Shielding			& $0.02 ~cm$ thick Tantalum strip, height $1.8 ~cm$, \\
				& around the collimator wall below the FOV to shield \\
			        & NaI+CsI crystal. Shielding weight $\sim 35 ~g$\\
\hline
\end{tabular}
\end{table}

\subsection{Summary and interpretation of the background simulation results}

In this section we present the energy response spectra (obtained 
from the simulation) in the scintillators NaI and CsI due to the various background 
components described in Section 2.2. These results enable us to see the contribution
of various individual background components to the total background.
In Figure 5a, we have shown the energy deposition spectrum in the NaI and CsI
crystals for the in-flight shielding configuration.  The NaI spectrum shows
a low energy cut-off below $\sim 15 ~keV$. Low energy photons are absorbed by 
the $0.5 ~mm$ Al filter used for protection of the crystals. The CsI
spectrum, also shows a few counts in the low energy range less than
$\sim 40 ~keV$. These counts are mainly due to the partial energy 
depositions of the higher energy photons. In higher energy range, the counts 
are high due to the photons coming through the collimator and not interacting
on the NaI crystal and as well as due to those photons entering into the
detector other than the collimator part. The peak in the NaI and CsI spectrum around
$60 ~keV$ is due to the Ta $K_{\alpha}$ fluorescent emission.
The detector response to the albedo Gamma-ray photons due to the Earth's 
atmosphere is given in Figure 5b. Here in the figure we can see the Ta 
$K_{\alpha}$ and $K_{\beta}$ fluorescent emission in the NaI spectrum along with the 
photon annihilation peak which is also visible in the CsI spectrum.
Figure 5c presents the energy deposition spectrum in the NaI and CsI 
crystal due to the downward going protons at low Earth orbit position.
We have shown the detector response spectrum in Figure 5d, for the 
upward going protons due to the interaction of the cosmic rays in the 
Earth's atmosphere. Figure 5e depicts the energy response spectrum in
the NaI and CsI crystals for the secondary neutron background spectrum.
From Figure 5(a,b,c,d,e) we can conclude that the most important contributor
to the prompt background noise are the CDGRB photons, while the secondary gamma-ray
photons also promote a significant portion of the noise.

\begin {figure}[htp]
\centering{
\includegraphics[width=0.49\textwidth]{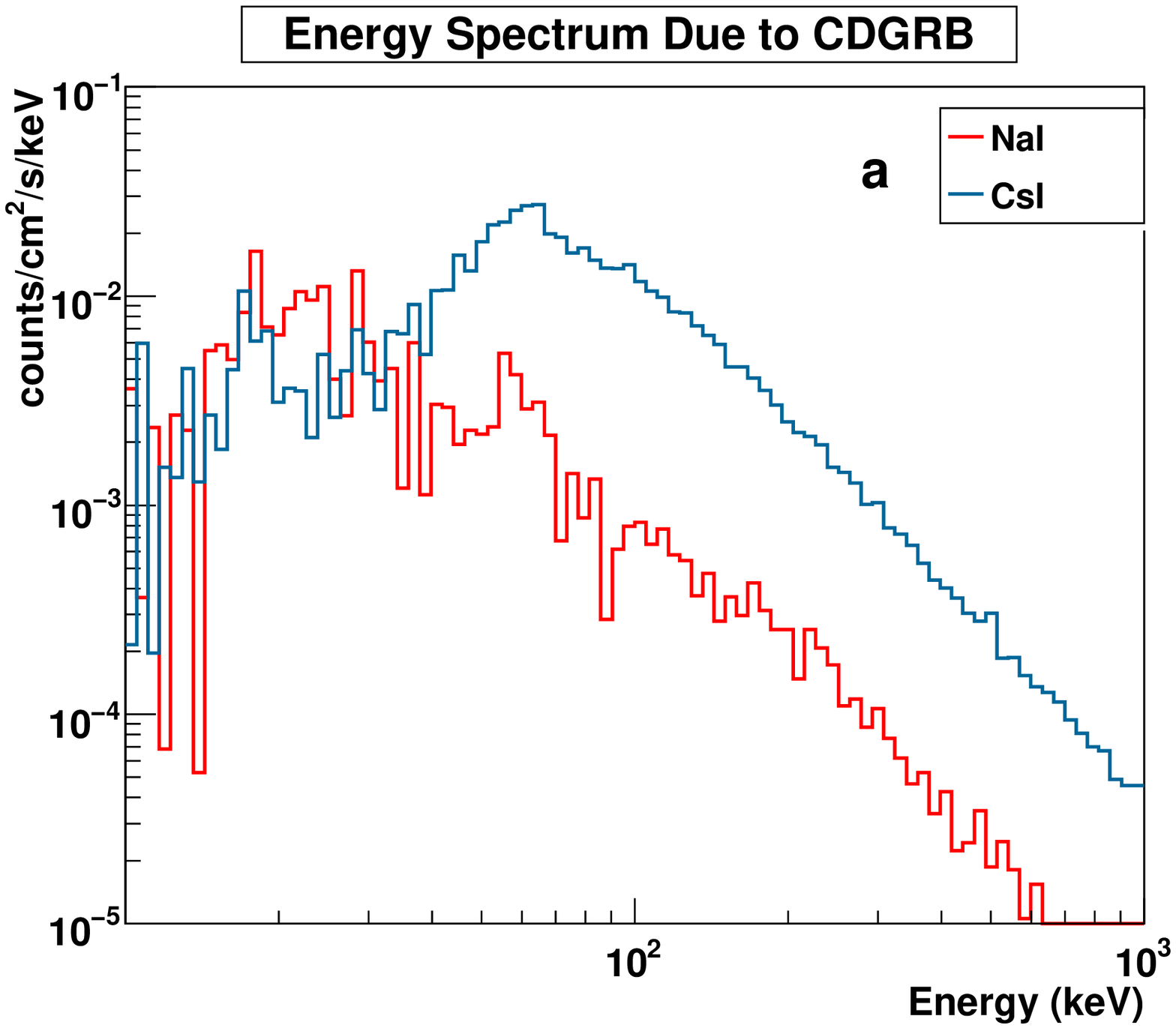}
\includegraphics[width=0.49\textwidth]{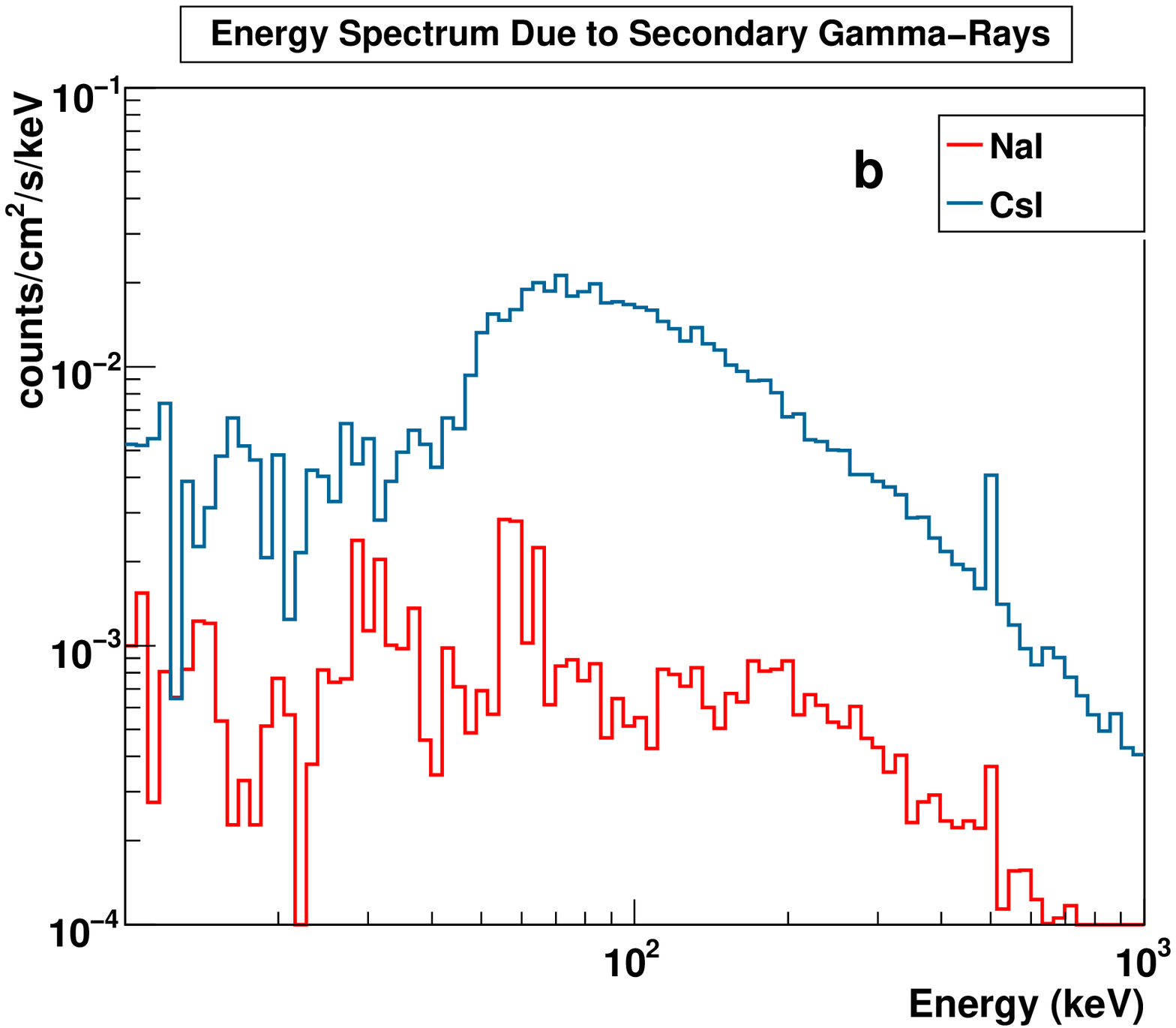}
\includegraphics[width=0.49\textwidth]{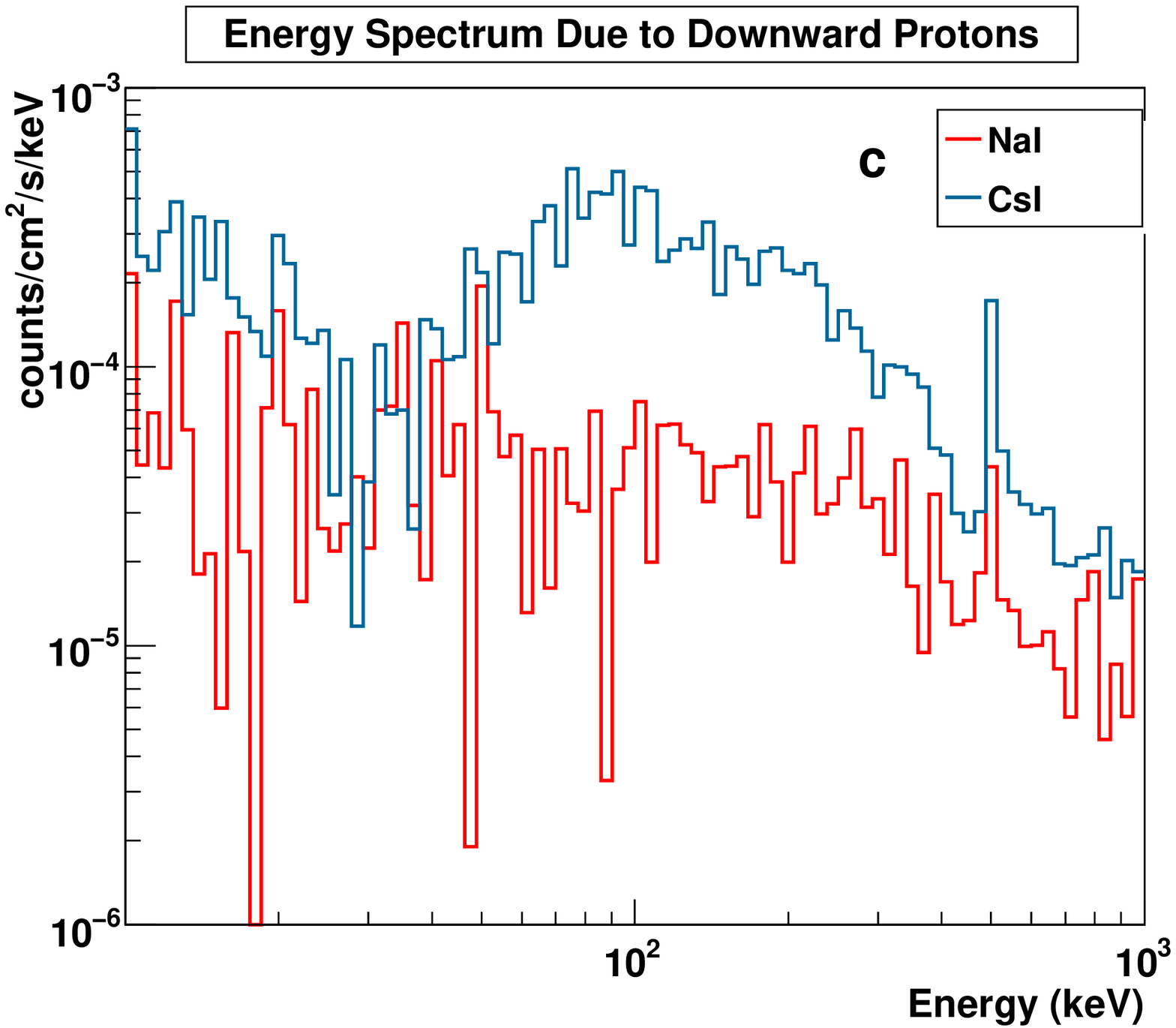}
\includegraphics[width=0.49\textwidth]{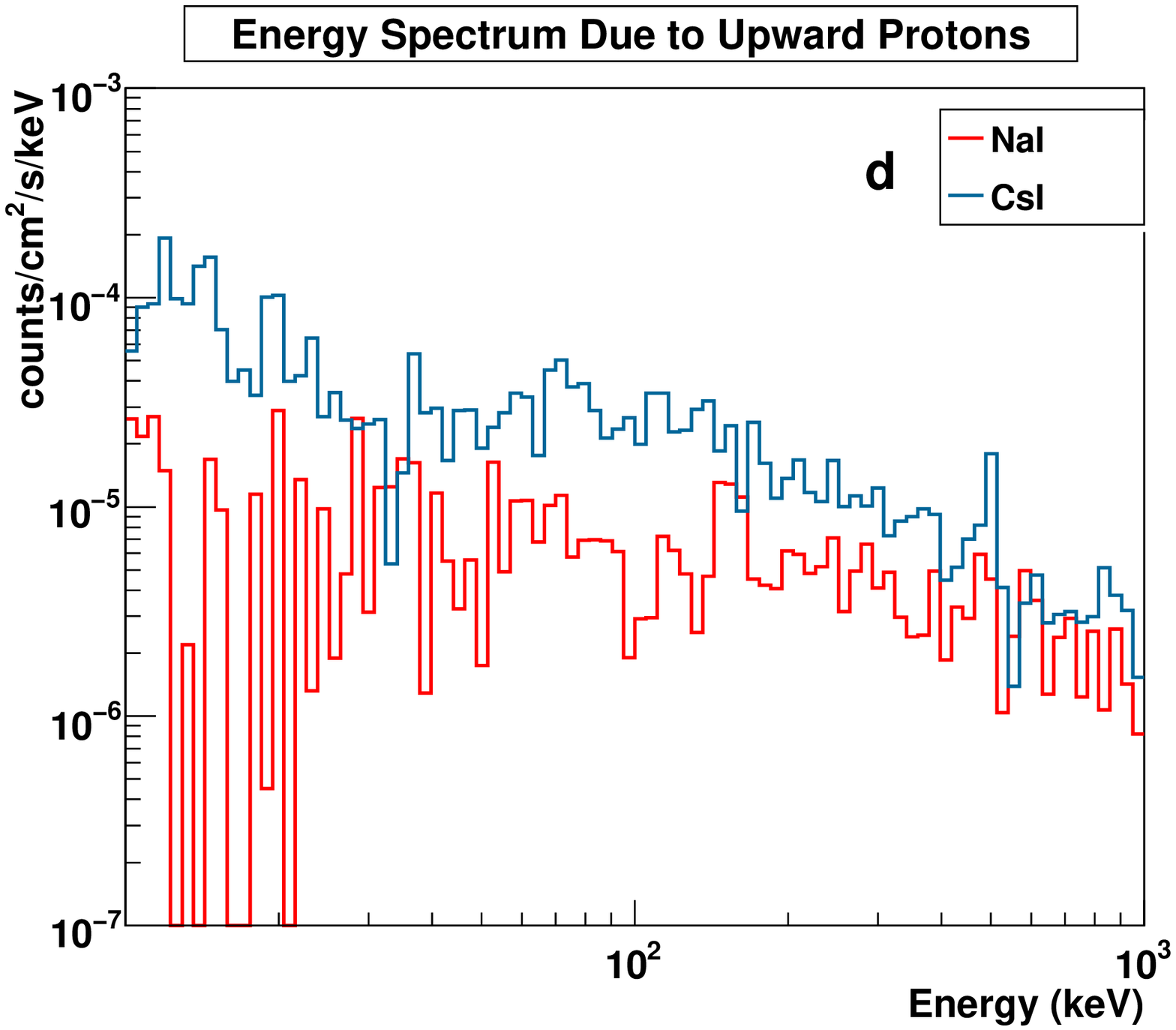}
\includegraphics[width=0.49\textwidth]{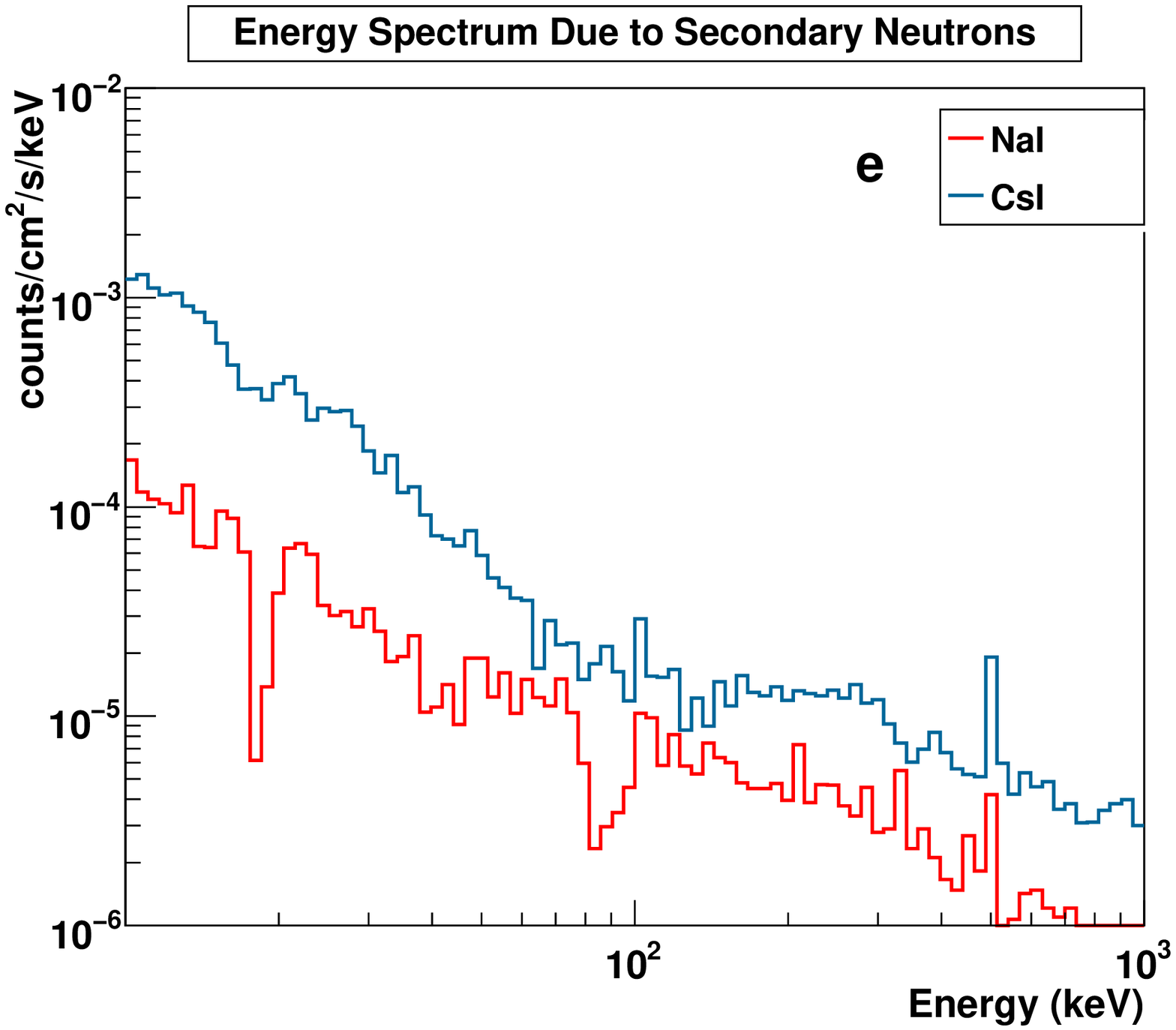}
\caption{The energy deposition spectrum in the NaI and CsI crystals due to
various background components incident spectrum of (a) CDGRB photon, 
(b) secondary gamma-ray photons, (c) downward going protons, (d) upward going 
protons in atmosphere, (e) albedo neutrons.}}
\end {figure}

The results from the simulation of RT-2/G are almost identical (except
the normalization in the low energy range) as that of the RT-2/S, since both of
the payloads have identical configuration except the FOV of the collimator
and $2 ~mm$ Al shielding on the top of the collimator of RT-2/G to 
have higher energy cut-off (below $25 ~keV$). So, the above results for
the RT-2/S are also applicable to RT-2/G payload regarding the 
response of the detector.

\subsection{Response of a GRB spectrum in RT-2/S}

The RT-2 detectors are primarily designed for spectroscopic measurements 
of solar flares in the $10 - 100 ~keV$ region. For efficient background 
rejection, a thick CsI detector was used.  While designing the experiment, 
it was realized that above $100 ~keV$, the RT-2 detectors will act as 
omni-directional gamma-ray burst detectors. Here we also simulate a GRB 
source and give the results below.

We compare the response of the detector for a known source 
over the total noise level due to various background components.
So we carry out a simulation for the incident GRB spectrum shown in Figure 3.
Moreover we consider two source positions: on-axis ($0^\circ$) and
off-axis ($50^\circ$) location with respect to the detector axis.

In Figure 6(a,b) we have shown the energy deposition spectrum in the NaI and
CsI crystals due to the GRB incident spectrum for the on-axis ($0^\circ$)
(in red colour) and off-axis ($50^\circ$) (in blue colour) source position.

\begin {figure}[htp]
\vbox{
\centering{
\includegraphics[width=0.49\textwidth]{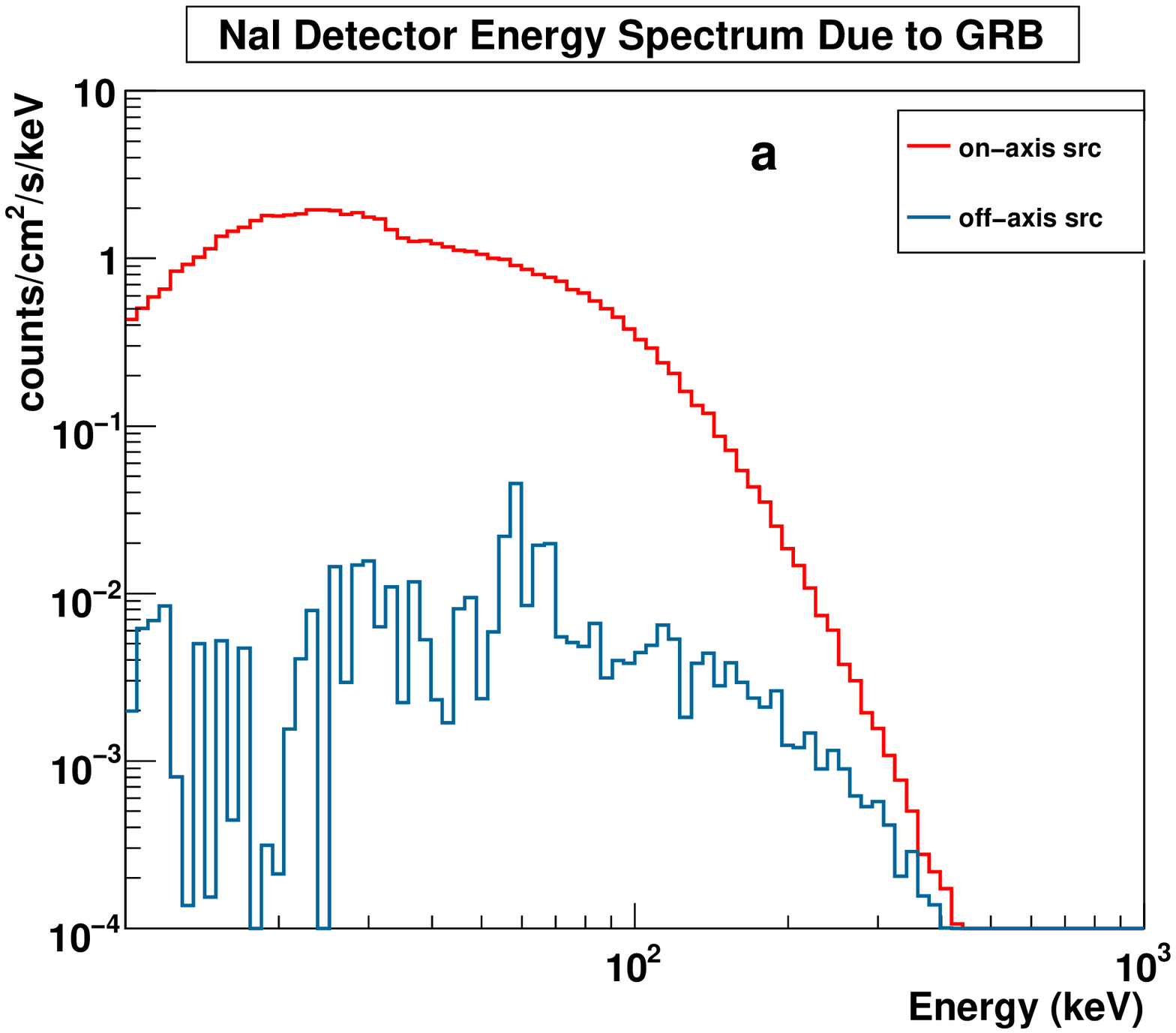}
\includegraphics[width=0.49\textwidth]{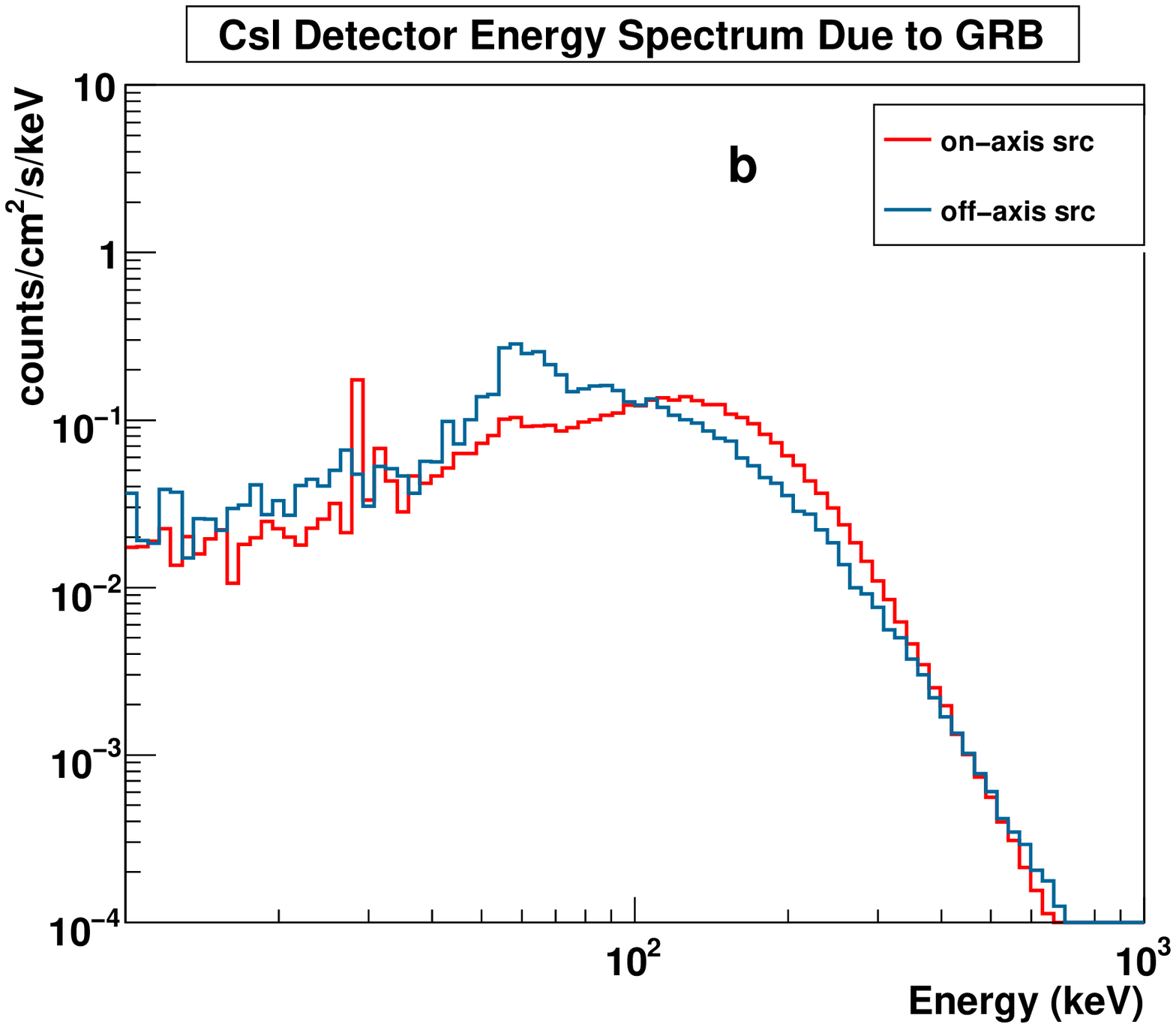}
\caption{The energy deposition spectrum in the (a) NaI and (b) CsI crystals
due the GRB incident spectrum (for on-axis ($0^\circ$) and off-axis ($50^\circ$) 
source position). }}}
\end {figure}

It is observed that the GRB spectrum in NaI is highly absorbed (due to the
FOV walls and shielding at the collimator wall) in the entire energy
range for off-axis position of the source (Figure 6a) and we can also see the
peak around $60 ~keV$ is due to the Ta $K_{\alpha}$ fluorescent emission.

In case of CsI crystal, the photon energy below $\sim100 ~keV$ is getting
absorbed both for off and on-axis source position. For on-axis, the photons
below $\sim100 ~keV$ is mostly detected by NaI crystal and we can also notice
a emission peaks around $30 ~keV$ due to the $K_{\alpha}$ and $K_{\beta}$
fluorescence of Iodine, whereas for off-axis the photons below $\sim100 ~keV$
are partially blocked by the shielding material around the lower part of the
collimator. Detection of source photon below $\sim25 ~keV$ is less significant
in CsI crystal. This feature is clearly seen for both the off and on-axis
source position (Figure 6b).

In Table 3, we present the number of photons from the incident GRB which deposit energy in
the NaI and CsI detectors. We have subdivided the whole energy range of
$10-1000 ~keV$ into smaller energy bands. We have given both the results for the on-axis
and the off-axis source positions and the counts due to the total {\it estimated} background 
(fourth column). The background counts within parenthesis is from the prompt sources. To obtain
the total backgrounds we need to include the contribution from the long-term
activations in the detector. In Section 5 we showed that this prompt 
background only accounts for up to $\sim 40\%$ of the total background in the NaI crystal.
So in order to compare the photon counts due to the GRB with the {\it total background}
noise, we have multiplied the background due to prompt emission
by a factor of $2.5$ and obtained the total background and presented in
in this column. This extra contribution, given that our satellite is polar, 
is reasonable since even for equatorial orbits the uncertainly could be 
a factor of 2 (Zoglauer, 2009).

\begin{table}[htp]
\noindent{Table 3: Photon counts in NaI and CsI Crystals in different energy
ranges due to the incident GRB spectrum for both the on-axis ($0^\circ$) and
off-axis ($50^\circ$) orientation of the source position and due to the
total {\it estimated} background spectrum with the simulated total prompt 
background in the bracket (see text for details).}
\vskip 0.5cm
\scriptsize
\centering
\begin{tabular}{|c|c|c|c|}
\hline
{\bf Energy range}     & {\bf on-axis ($0^\circ$)} & {\bf off-axis ($50^\circ$)} & {\bf total bkg. (prompt)} \\
{\bf in (keV)}     & {\bf GRB} & {\bf GRB} & \\
\cline{2-4}
                   & \multicolumn{3}{|c|}{\bf NaI} \\
\hline
$10-20$ & $1348.9$ & $2.9$ & $17.0 (6.8)$ \\
\hline
$20-50$ & $4764.7$ & $21.8$ & $45.5 (18.2)$ \\
\hline
$50-100$ & $3449.5$ & $54.3$ & $37.2 (14.9)$ \\
\hline
$100-150$ & $996.4$ & $22.3$ & $16.8 (6.7)$ \\
\hline
$150-1000$ & $296.3$ & $26.9$ & $70.3 (28.1)$ \\
\hline
 & \multicolumn{3}{|c|}{\bf CsI} \\
\hline
$10-20$ & $21.1$ & $32.0$ & $25.8 (10.3)$ \\
\hline
$20-50$ & $167.0$ & $210.3$ & $118.1 (47.2)$ \\
\hline
$50-100$ & $512.1$ & $953.3$ & $474.6 (189.8)$ \\
\hline
$100-150$ & $692.5$ & $552.1$ & $297.9 (119.2)$ \\
\hline
$150-500$ & $848.5$ & $516.6$ & $533.6 (213.4)$ \\
\hline
$500-1000$ & $4.8$ & $6.2$ & $110.3 (44.1)$ \\
\hline
\end{tabular}
\end{table}

From the results of Table 3 we can see that for the spectrum of GRBs we are
interested, NaI has a quite large source to noise ratio (S/N) in the energy range $10-150 ~keV$ for on-axis case. However, for the off-axis case, the S/N does not permit to detect the 
source at all. On the other hand, for the on-axis case, 
CsI has a quite high S/N value in the energy range
of $50-500 ~keV$ and for the off-axis case this energy range gets broadened 
in $20-500 ~keV$. So this simulation of the GRB ensures the capability of 
such source detection by the RT-2/S instrument. Based on the 
orbital and detector configurations a conservative estimate suggests 
a detection of $\sim 20$ GRBs per year.

\section{Simulation of RT-2/CZT payload}

The parameters used to construct the virtual detector of RT-2/CZT payload 
within the Geant4 toolkit environment is given in Table 4. These parameter
values are same as that used in onboard detector design and we applied some
simplifications as we have already mentioned in section 3. A 3D view of
the RT-2/CZT payload along with different components (e.g., detectors, CAM (Coded 
Aperture Mask), FZP (Fresnel Zone Plate), collimator etc.) as considered for 
the simulation is shown in Figure 7. This figure is the direct output from 
the simulation.

\begin {figure}[htp]
\centering{
\includegraphics[width=8.0cm]{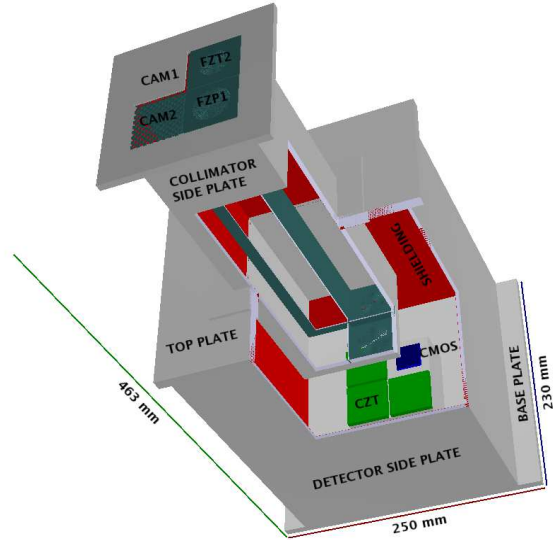}
\caption{A schematic 3D view of RT-2/CZT payload with different components
(eg. detectors, CAM (Coded Aperture Mask), FZP (Fresnel Zone Plate), collimator 
etc.).}}
\end {figure}

\begin{table}[htp]
\noindent{Table 4: Payload construction parameter specifications of RT-2/CZT.}
\vskip 0.5cm
\scriptsize
\centering
\begin{tabular}{|l|c|}
\hline
{\bf Detector Parts} 		& {\bf Dimensions} \\
\hline
Bottom plate 			& $25.0\times23.0\times0.8~cm^3$ \\
\hline
Detector Housing 		& $20.2\times22.4\times26.7~cm^3$ (inner dimension)\\
                 		& wall thickness $0.3~cm$ \\
\hline
Top Plate 			& $23.0\times23.0\times1.0~cm^3$ \\
\hline
External Collimator 		& $10.7\times10.7\times17.8~cm^3$ (inner dimension)\\
                    		& wall thickness $0.4~cm$ \\
\hline
Internal Collimator 		& $10.7\times10.7\times13.0~cm^3$ (inner dimension)\\
                    		& wall thickness $0.4~cm$ \\
\hline
FOV 				& $0.1~cm$ Aluminum shielded by $0.007~cm$ Tantalum \\
    				& two perpendicular walls to divide the collimator in four quadrants. \\
\hline
CZT Crystal 			& $4.0\times4.0\times0.5~cm^3$ (3 crystals) \\
            			& gap between modules $0.25~cm$ \\
\hline
CMOS 				& $2.4\times2.4\times0.3~cm^3$ \\
\hline
CAM 				& $0.1~cm$ thick Tantalum CAM on top of two collimator quadrants. \\
    				& $1 ~mm$ Al filter used for one CAM and other one is open to sky. \\
\hline
Zone Plate 			& $0.1~cm$ thick Ta zone plates at the face and end \\
           			& of the collimator on one CZT and CMOS crystals \\
\hline
Shielding 			& $0.007 ~cm$ Tantalum around the collimator wall. \\
				& $0.02 ~cm$ Tantalum below the top plate. \\
				& $0.02 ~cm$ Tantalum around the detector housing side plates. \\
				& Shielding weight $\sim 951 ~g$. \\
\hline
\end{tabular}
\end{table}

\subsection{Summary and interpretation of the background simulation results}

In the same way and for the same purpose as we have discussed for NaI and
CsI crystals in RT-2/S instrument, we have calculated the energy response 
spectrum for the three CZT and one CMOS detectors in RT-2/CZT.
In Figure 8a, we have depicted the energy deposition spectrum in one of
the CZT and CMOS due to the CDGRB spectrum. For the CZT spectrum,
we can see a larger amount of fluorescent photons (emission peak at
$\sim60 ~keV$) due to Ta shielding of thickness $0.2 ~mm$ on top plate
and around the side plates of the payload. The simulated results for the other
two CZT modules are mostly identical and hence not discussed here.
The energy response spectrum in CZT and CMOS due to the secondary gamma-ray
spectrum is shown in Figure 8b. These spectra also show more or less the same 
features as the primary gamma-ray (CDGRB) spectrum. Figure 8c presents the energy 
deposition spectrum in the two crystals for the downward going protons,
while Figure 8d shows the same for the upward going protons. In these two 
cases we can see a relatively higher energy deposition in the CMOS than in the CZT.
The energy response of CZT and CMOS for the secondary neutron spectrum is given in Figure 8e.
Figure 8(a-e) also shows the dominance of the CDGRB and secondary photon spectrum 
in the noise contribution.

\begin {figure}[htp]
\centering{
\includegraphics[width=0.49\textwidth]{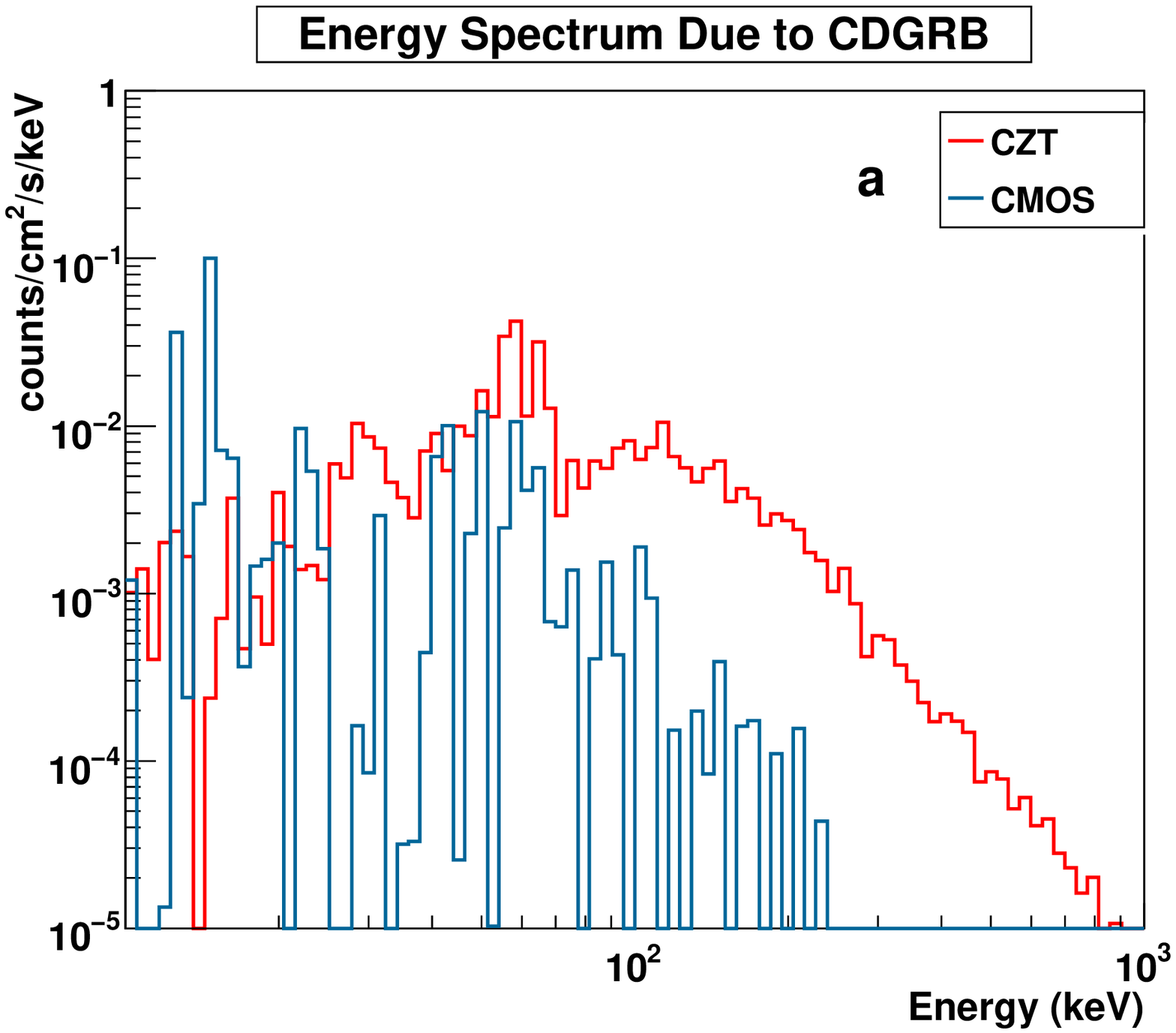}
\includegraphics[width=0.49\textwidth]{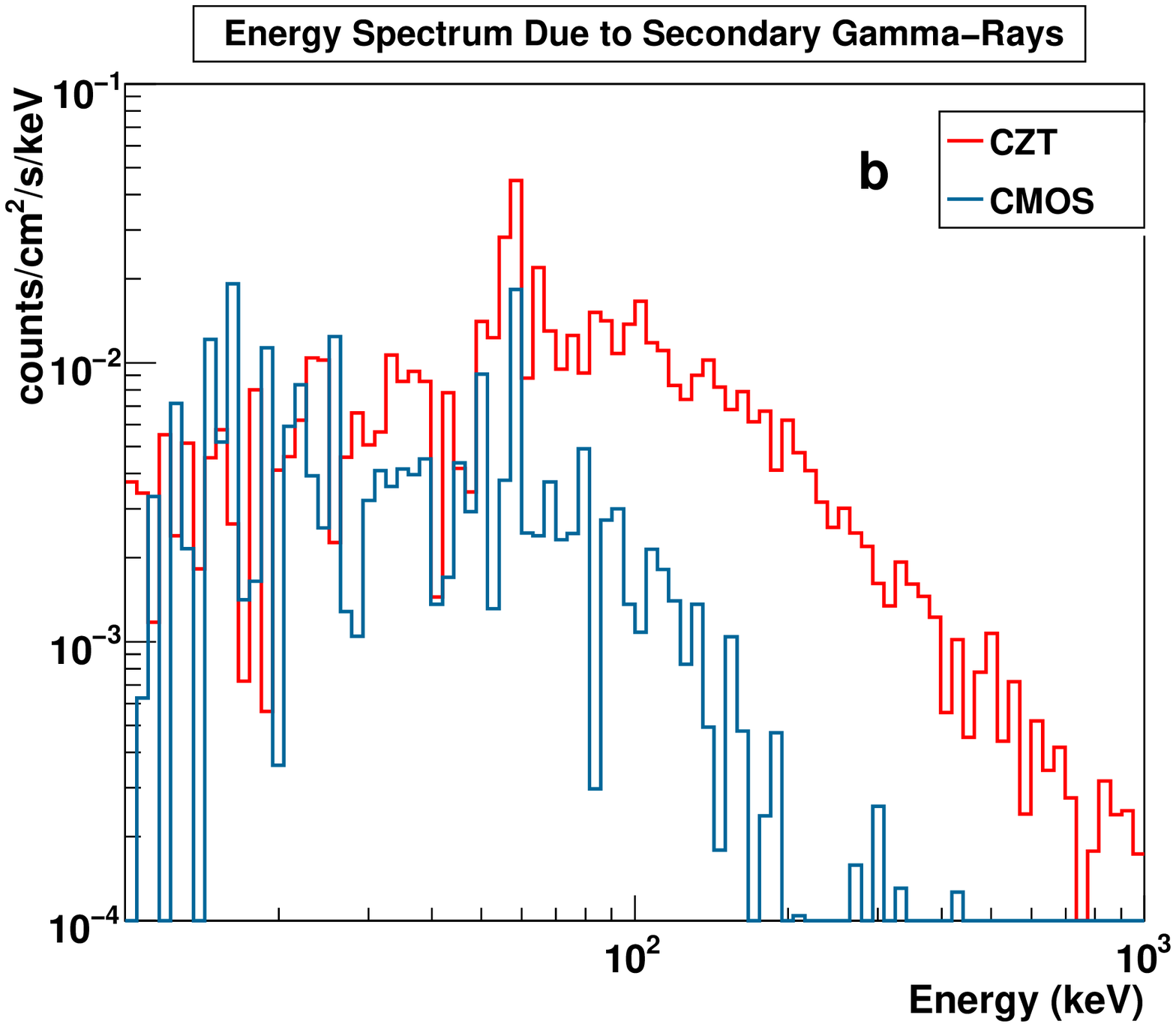}
\includegraphics[width=0.49\textwidth]{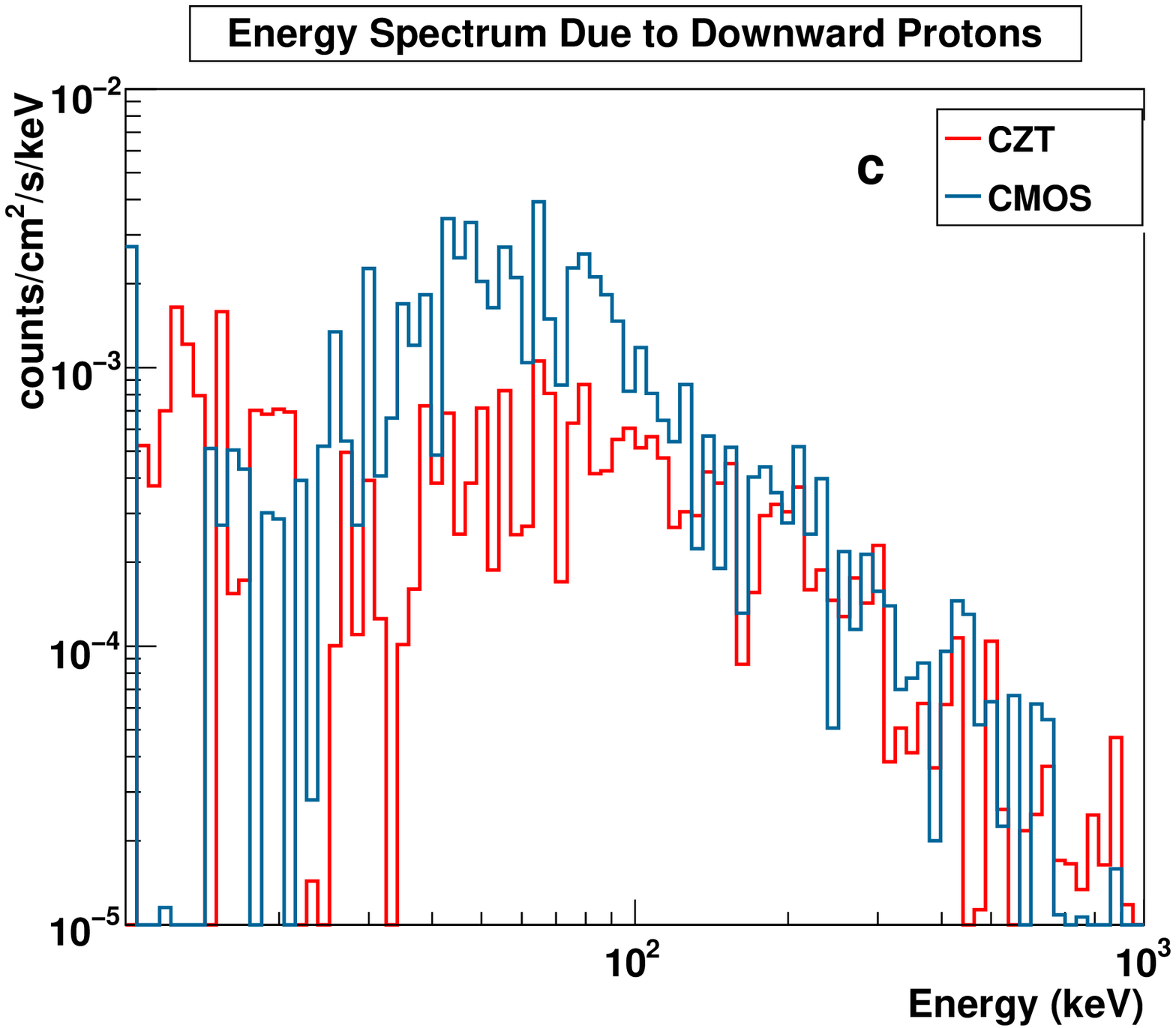}
\includegraphics[width=0.49\textwidth]{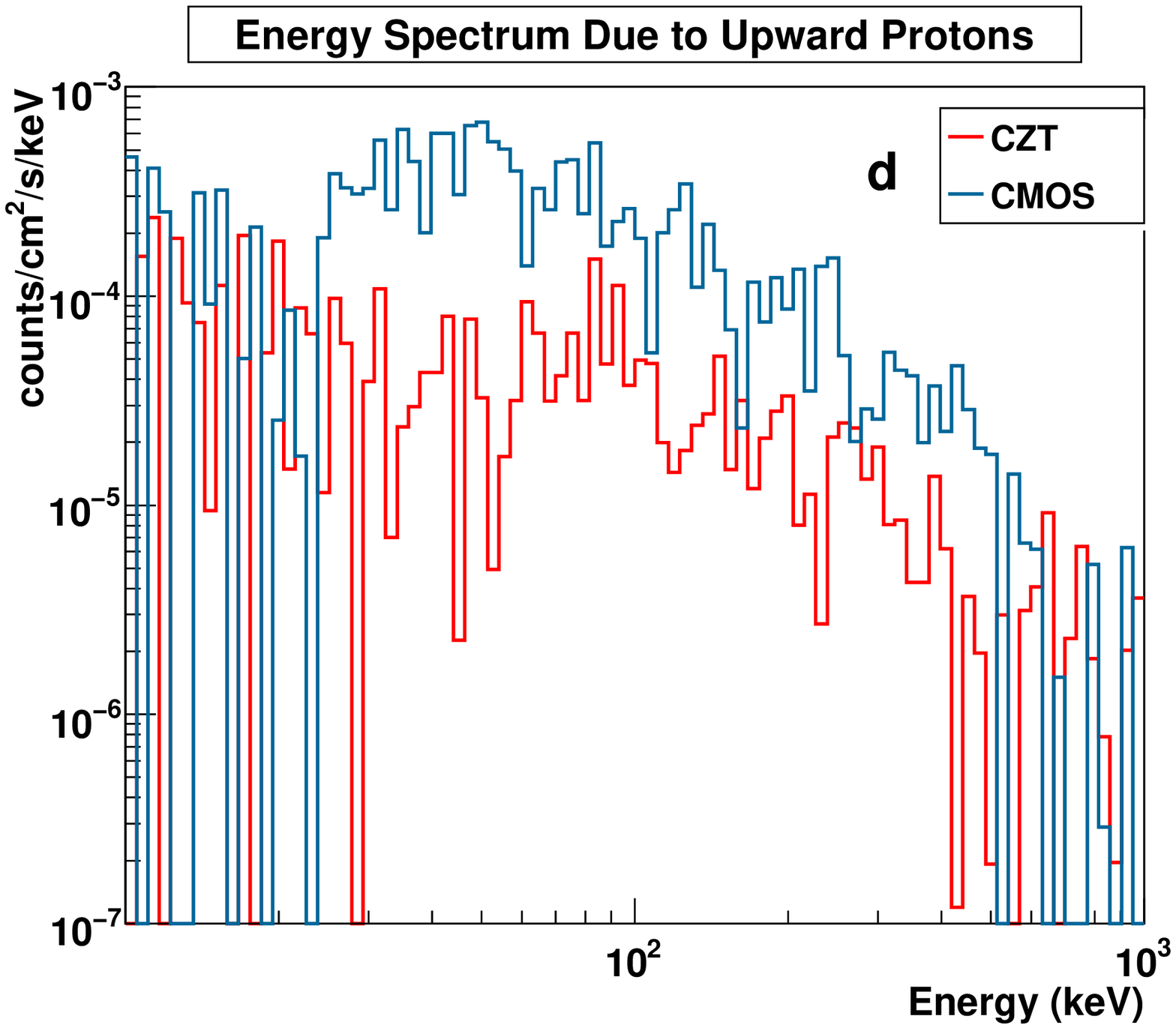}
\includegraphics[width=0.49\textwidth]{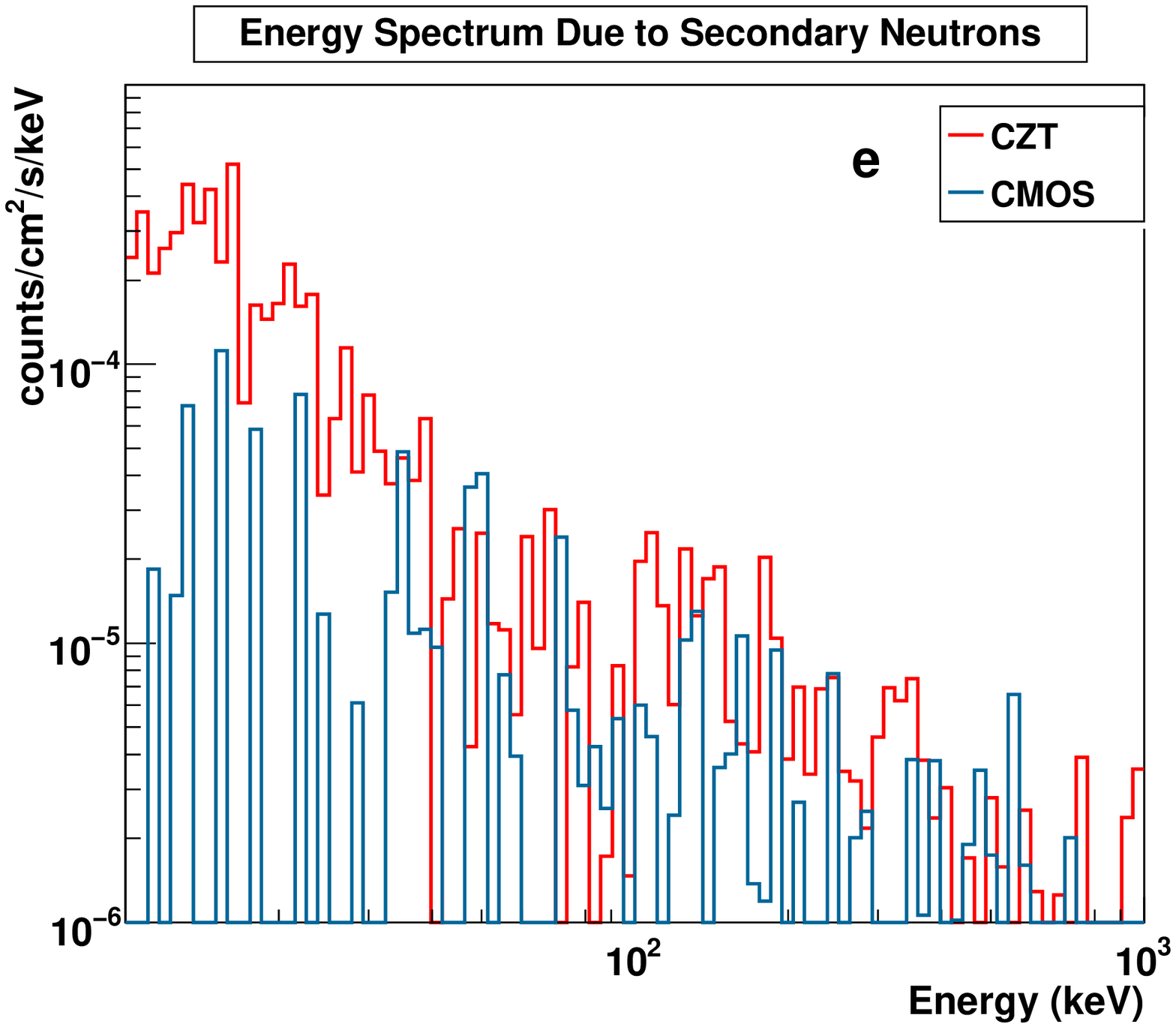}
\caption{The energy deposition spectrum in the CZT and CMOS due 
to (a) CDGRB photons, (b) secondary gamma-ray photons, (c) downward going protons,
(d) upward going protons and (e) albedo neutrons. }}
\end {figure}

\subsection{Response of a GRB spectrum in RT-2/CZT}

We now simulate the detector for a typical GRB spectrum (Eqn. 7) as the
incident spectrum shown in Figure 3 for two source location as mentioned
in section 3.2.

\begin {figure}[htp]
\vbox{
\centering{
\includegraphics[width=0.49\textwidth]{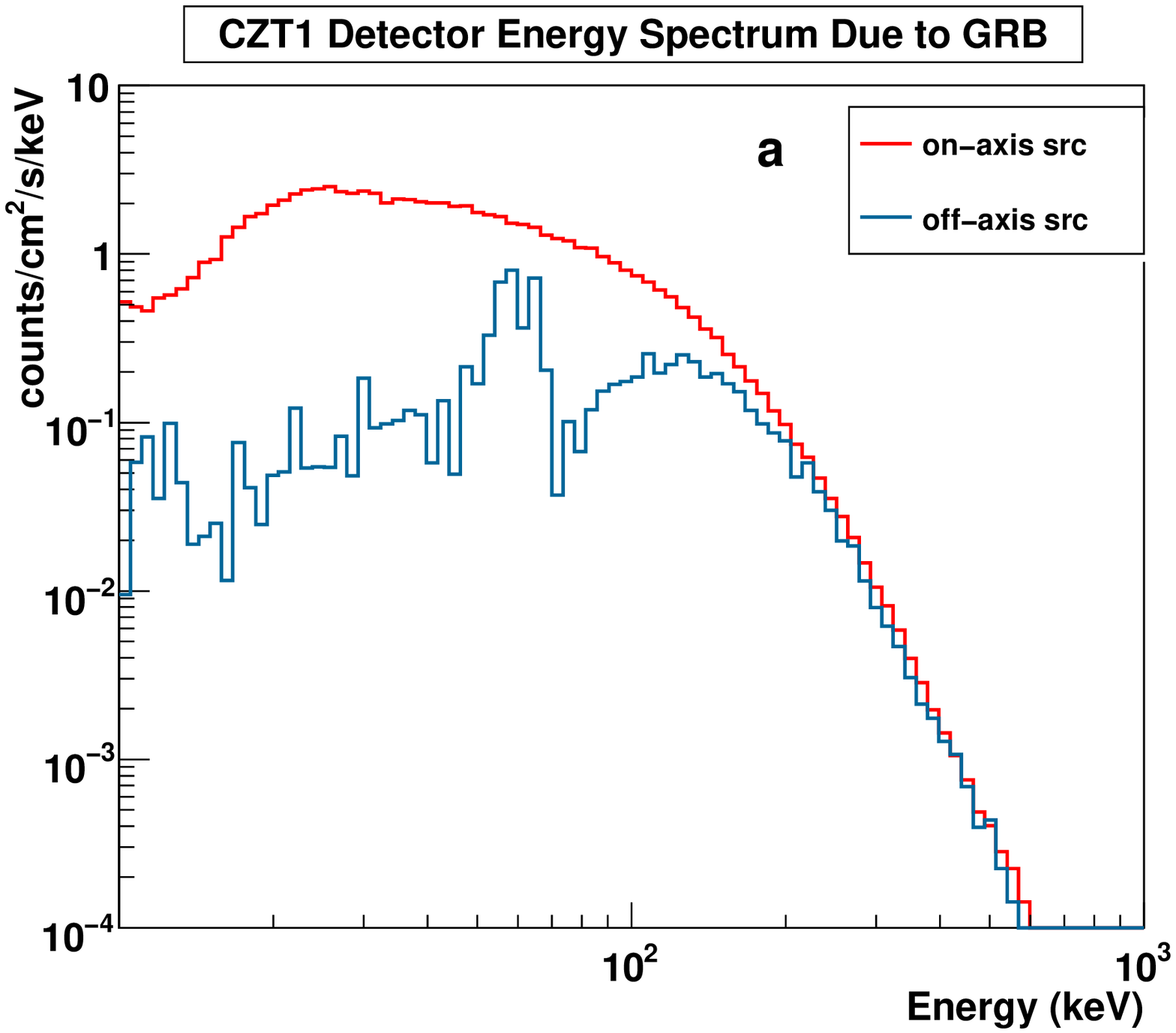}
\includegraphics[width=0.49\textwidth]{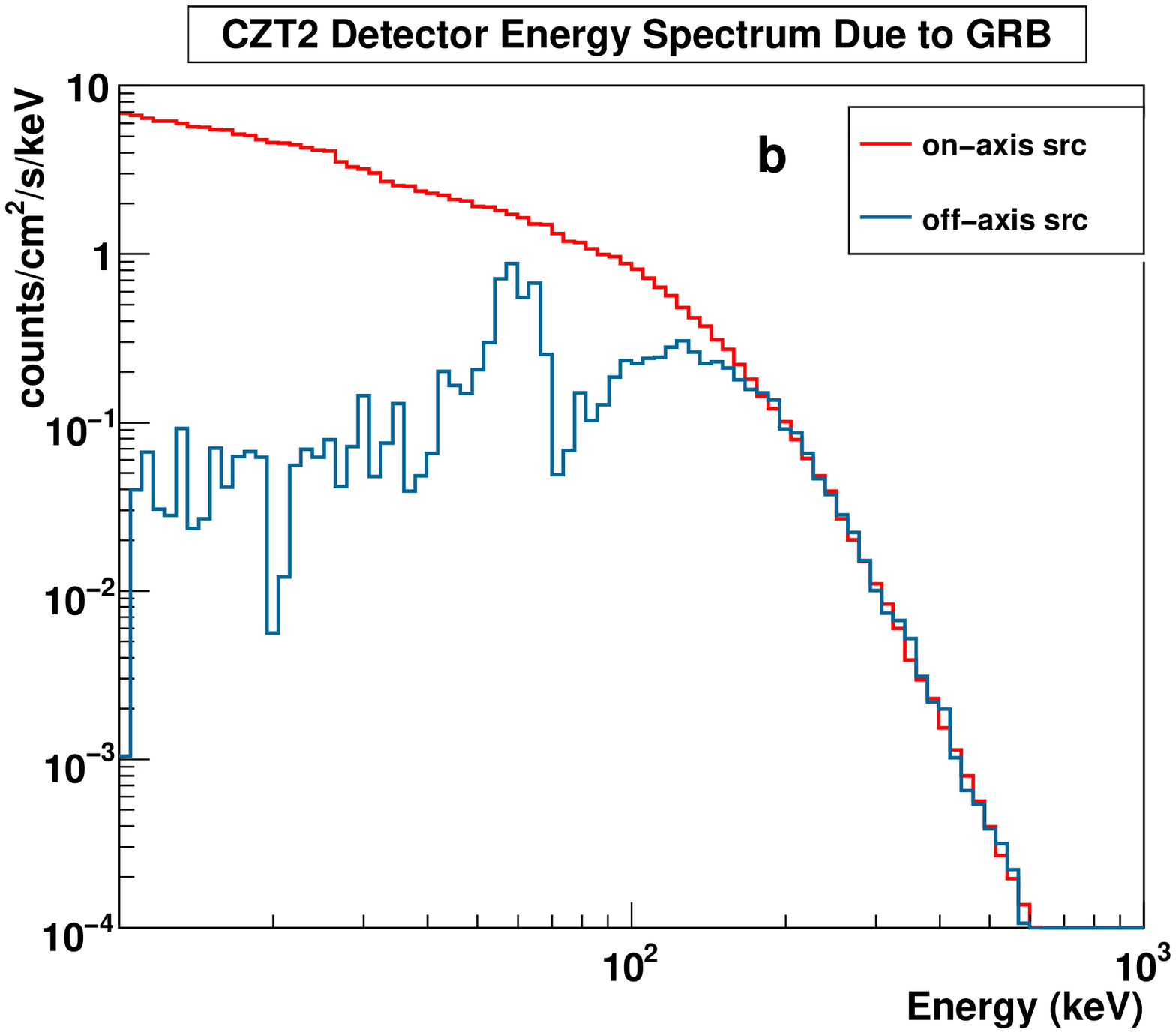}
\includegraphics[width=0.49\textwidth]{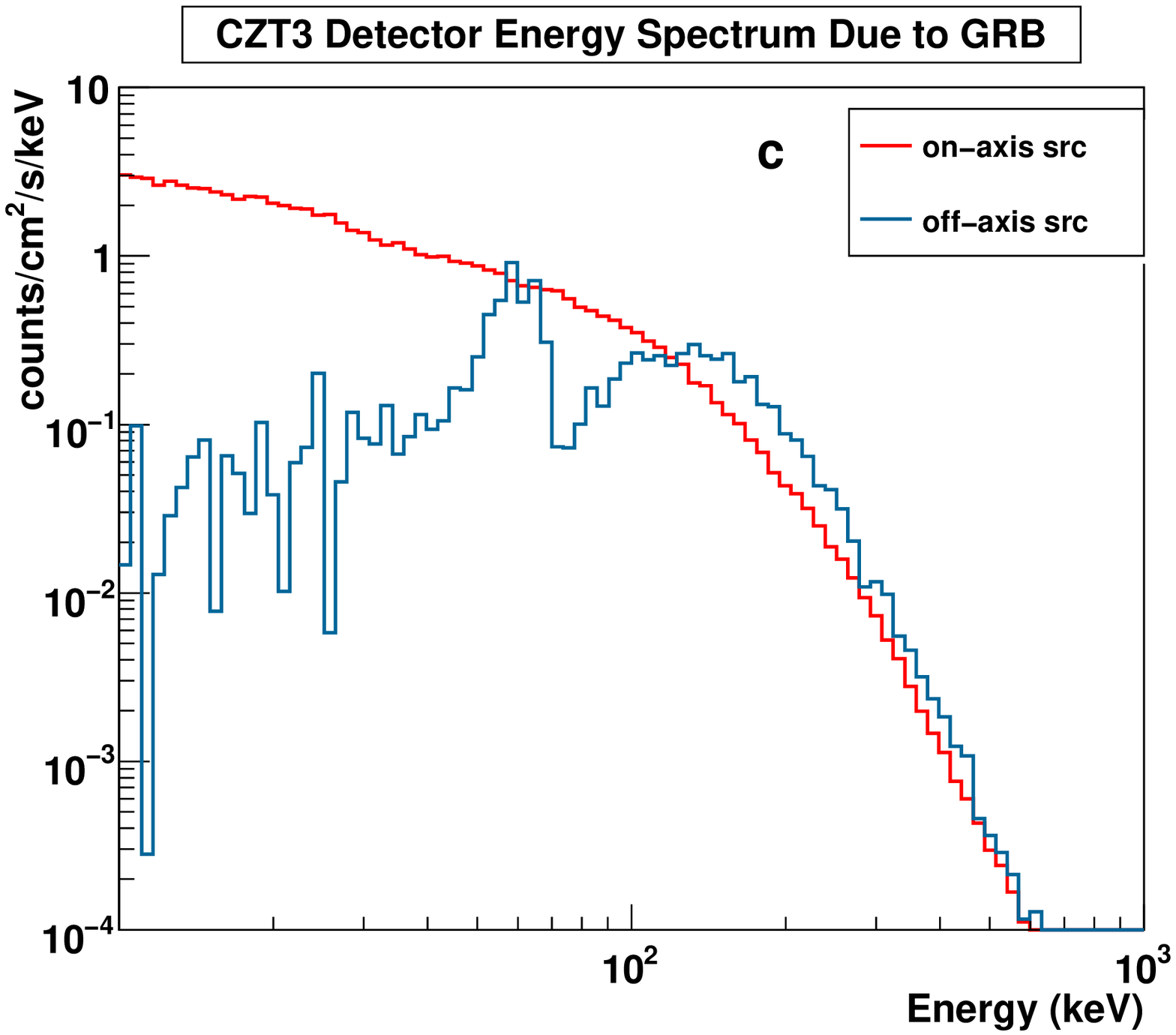}
\includegraphics[width=0.49\textwidth]{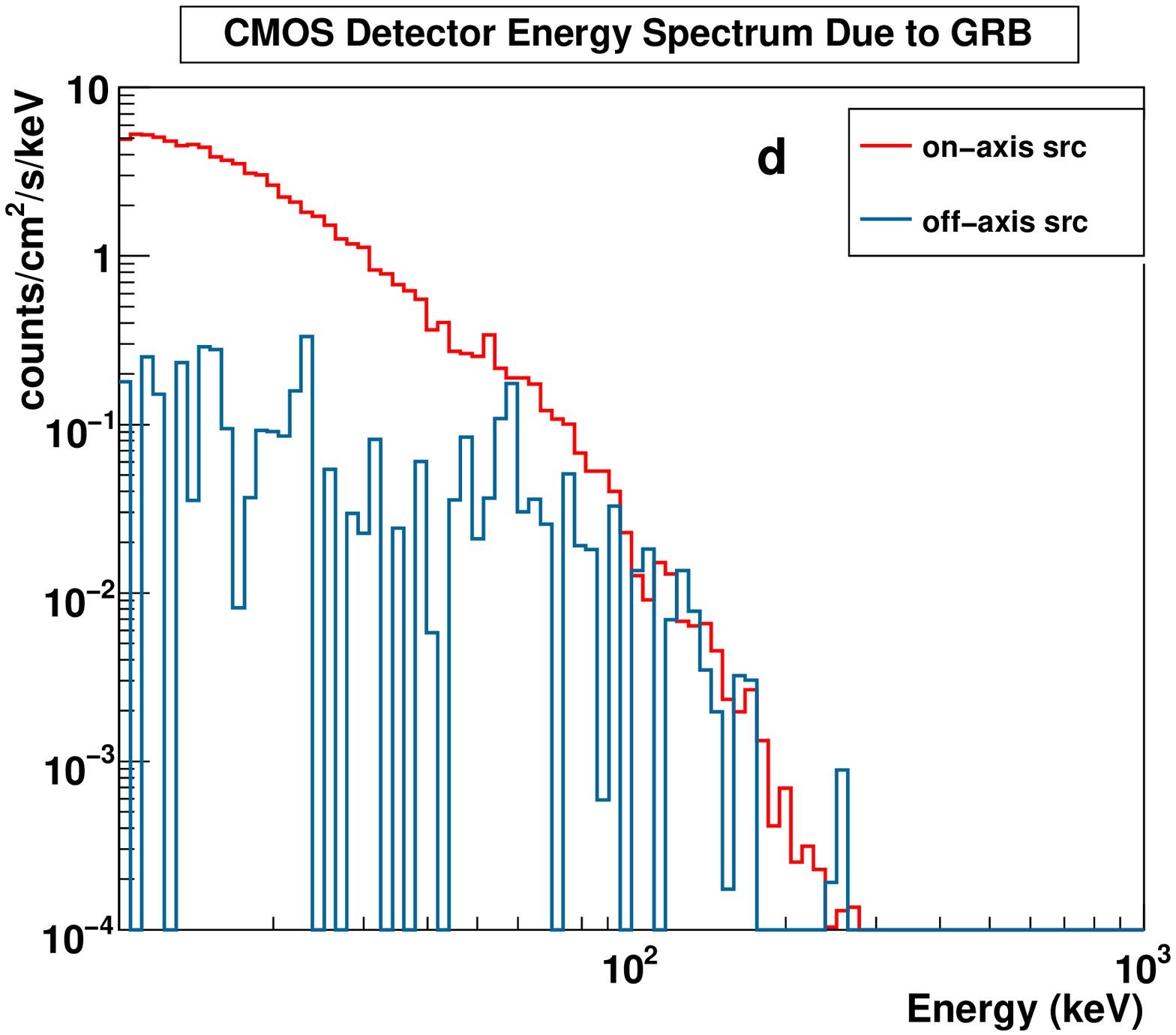}
\caption{The energy deposition spectrum in the CZT and CMOS crystals due the
GRB incident spectrum for on-axis ($0^\circ$) and off-axis ($50^\circ$)
source positions.
(a) Spectrum in CZT with a CAM along with Al sheet, (b) that in CZT with a
CAM, (c) that in CZT with a pair of zone plates and (d) that in CMOS with a
pair of zone plates.}}}
\end {figure}

In Figure 9(a-d), we plot the energy deposition spectrum
separately for the three CZT modules and CMOS detector both for the on-axis
($0^\circ$) and off-axis ($50^\circ$) GRB position. Figure 9a shows the energy
deposition spectrum in the CZT (CZT1) containing a CAM along with an $1.0 ~mm$
thick Al filter in front of the collimator facing the sky. The
spectrum in the CZT (CZT2) containing a CAM at the face of the collimator is 
given in Figure 9b.
Figure 9c presents the same in the CZT (CZT3) containing a pair of aligned zone
plates (FZPs) placed at $32 ~cm$ apart in third quadrant of the collimator.
In Figure 9d we have shown the spectrum of CMOS detector which is also having 
two aligned zone plates (FZPs) placed in the fourth quadrant of collimator. It is evident 
from Figure 9(a-c) that all the CZT spectra are roughly the same irrespective of
their different configurations (Nandi et al. 2010) of the collimator for the
off-axis ($50^\circ$) source position though they differ for the on-axis source
position (low energy photons are absorbed in CZT1 due to $1 ~mm$ Al shielding).
This is because for the off-axis source position most of the photons which are
depositing their energy into CZT or CMOS are coming through the detector side
plates having uniform shielding material. Those photons interacting on the mask
pattern are less likely to go to the CZT and CMOS to deposit their energy. All
the CZT spectra show a prominent Ta fluorescent peak (around $60 ~keV$), caused
by the shielding material.

In Table 5, we give the number of photon counts for three CZT modules and for
CMOS detector due to the GRB spectrum for the on-axis ($0^\circ$) and off-axis
($50^\circ$) source positions and due to total {\it estimated} background spectrum
along with the prompt background noise as discussed in section 3.
We have subdivided the total energy range of $10-1000 ~keV$ into several
energy bands. These results could be useful while analyzing the data from real
observation (GRB) with RT-2 instruments.

\begin{table}[htp]
\noindent{Table 5: Photon counts in the three CZTs and CMOS detector in
different energy ranges due to the GRB spectrum for both on-axis ($0^\circ$)
and off-axis ($50^\circ$) source position and due to total estimated background 
spectrum with the simulated total prompt background inside the bracket (see text 
for detail).}
\vskip 0.5cm
\scriptsize
\centering
\begin{tabular}{|c|c|c|c|c|}
\hline
{\bf Energy range} & {\bf CZT1} & {\bf CZT2} & {\bf CZT3} & {\bf CMOS}\\
\cline{2-5}
{\bf (in keV)} & \multicolumn{4}{|c|}{\bf on-axis ($0^\circ$) GRB} \\
\hline
$10-20$ & $182.1$ & $936.3$ & $414.3$ & $242.9$\\
\hline
$20-50$ & $1049.0$ & $1415.0$ & $617.5$ & \\
\cline{1-4}
$50-100$ & $939.8$ & $1009.8$ & $438.3$ & $184.4$ \\
\cline{1-4}
$100-150$ & $408.5$ & $422.6$ & $186.7$ & \\
\hline
$150-200$ & $141.3$ & $145.5$ & $64.4$ & $0.6$ \\
\cline{1-4} 
$200-1000$ & $64.7$ & $66.6$ & $36.9$ & \\
\hline
 & \multicolumn{4}{|c|}{\bf off-axis ($50^\circ$) GRB} \\
\hline
$10-20$ & $7.0$ & $7.5$ & $7.9$ & $7.2$\\
\hline
$20-50$ & $52.8$ & $51.0$ & $55.7$ & \\
\cline{1-4}
$50-100$ & $210.2$ & $230.8$ & $238.4$ & $20.8$ \\
\cline{1-4}
$100-150$ & $173.5$ & $203.5$ & $207.8$ & \\
\hline
$150-200$ & $98.9$ & $130.5$ & $137.6$ & $0.4$ \\
\cline{1-4}
$200-1000$ & $51.7$ & $68.9$ & $68.0$ & \\
\hline
& \multicolumn{4}{|c|}{\bf total bkg. (prompt)} \\
\hline
$10-20$ & $2.8 (1.1)$ & $2.5 (1.0)$ & $4.3 (1.7)$ & $2.5 (1.0)$\\
\hline
$20-50$ & $16.2 (6.5)$ & $17.7 (7.1)$ & $16.3 (6.5)$ & \\
\cline{1-4}
$50-100$ & $59.3 (23.7)$ & $56.5 (22.6)$ & $59.3 (23.7)$ & $11.1 (4.5)$ \\
\cline{1-4}
$100-150$ & $33.2 (14.1)$ & $34.5 (13.8)$ & $35.9 (14.4)$ & \\
\hline
$150-200$ & $24.5 (9.8)$ & $21.2 (8.5)$ & $23.2 (9.3)$ & $1.9 (0.8)$ \\
\cline{1-4}
$200-1000$ & $34.6 (13.9)$ & $35.8 (14.3)$ & $39.1 (15.6)$ & \\
\hline
\end{tabular}
\end{table}

From the results in Table 5 it appears that with CZT detector, we can detect GRBs 
quite confidently in the energy range of $10-200$keV for the on-axis case and $50-200 ~keV$ 
for the off-axis case. 

\section{Comparison with Observed Data}

We compare the simulation results for the various background components with
the real data measured by the detector in its in-flight operation. We consider
one set of onboard NaI spectral data of RT-2/S near the equatorial region and far from the 
trapped charged particle regions. This data refers to the satellite position 
when the instruments came out of the polar region, the count rates were found to
be dropping slowly. We noticed that the low energy count rates stabilized
faster than the high energy count rates. We took the background region
near the equatorial region when: (a) the low energy ($<100 ~keV$) count rates 
were steady (better than a percent stability in $10$ minutes) and (b) when the 
high energy count rates ($>330 ~keV$) were steady by about $5\%$ in $10$ minutes.

The primary energy range of the instrument for spectroscopic measurements is 
$\sim15 - 100 ~keV$ in the NaI detector. Hence we have concentrated on predicting 
background around this region.
 
The comparison of the data in the energy range of $20-110 ~keV$ with the simulated background 
components of primary and secondary gamma rays, downward and upward going protons 
(consisting primary CR and secondary protons) and secondary neutrons
are shown in Figure 10. 

\begin {figure}[htp]
\centering{
\includegraphics[width=9cm]{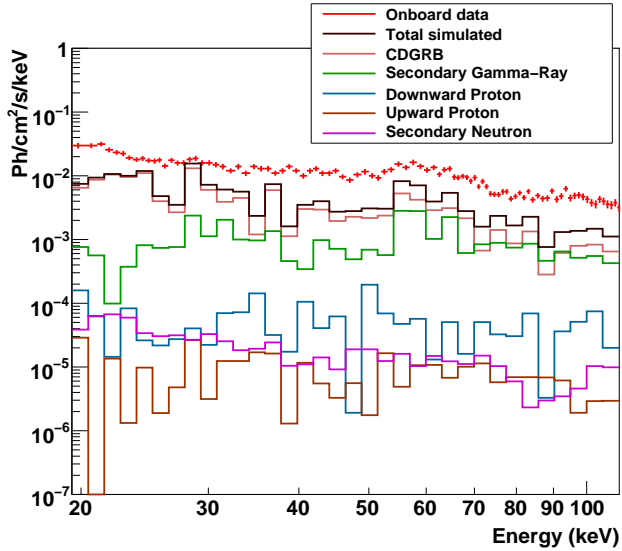}
\caption{ Onboard RT-2/S NaI background spectrum along with various simulated 
background components of CDGRB photons, secondary photons, downward going protons,
upward going protons and albedo neutrons. 
}}
\end {figure}

From Figure 10 we observe that the total simulated background noise due to
the main sources of the prompt background components are not sufficient
to explain the total measured background flux. This prompt background components
are able to address only about $40\%$ of the measured background. The rest of 
the background noise is suspected to be due to the long-term detector 
activation which is to be considered in a future work. 
Among the simulated components the most significant contributor ($\sim 27\%$) 
to the background noise in the NaI crystal are the CDGRB photons.
albedo photons contribute $\sim 10\%$, downward going protons $\sim 0.5\%$ and 
rest from the other factors.

For onboard calibration of the detectors (RT-2/S and RT-2/G), we used radioactive source 
Co-57 (122 keV) that is placed into one of the slats of the collimator. Extensive 
calibration of both the Phoswich detectors (NaI/CsI) on the ground reveals that 
the $\sim58 ~keV$ emission line (Debnath et al. 2010) is the intrinsic detector 
background feature (not due to the Tantalum (Ta) shielding around the collimator mesh). 
The emission feature of $\sim 58 ~keV$, earlier reported by Gruber et al. (1996) 
was also observed in the onboard data of RT-2/S and RT-2/G (Nandi et al. 2009).
The extra emission line feature at around $20 ~keV$ (cutoff energy of RT-2/S 
is around $15 ~keV$) could be due to the detector's electronic noise, which was
not observed during ground testing and the investigation is still ongoing. These two emission
features characteristics will be taken up in a separate paper, while considering the
long-term activation aspects of the background components in the detector.
Also the comparison of the simulated result with the real data from the 
other active volumes are under investigation due to the calibration concern
of the detectors.

The CsI detector in the phoswich combination is primarily used for background 
rejection and we use it only for very bright sources like GRBs. To investigate 
the background in CsI for faint sources, one also needs to simulate the exact 
instrument characteristics like the pulse shape for partial energy deposition 
in NaI and CsI. These activities would be taken up in a separate paper. We have, 
however, looked at the CsI spectra and ensured that the background spectral shape 
broadly agrees with the model prediction.

\section{Discussions and Conclusions}

Background simulations of space-borne payloads is one of the challenging tasks
to understand the space environment as well as the effect of high energy
radiation (photons, charged particles, neutrons) on the detectors itself. 
The effects of the major prompt background components like cosmic diffused gamma-ray 
background, secondary gamma-ray photons, primary cosmic-ray protons, secondary protons 
and albedo neutrons on the RT-2 payloads (RT-2/S, RT-2/G and RT-2/CZT) are studied in detail,
which helped the background calibration and source data extraction from the 
RT-2 Experiment. The weight of the material (Ta) that is used for
shielding purpose for RT-2/S (RT-2/G) and RT-2/CZT payload is around $35 ~g$
and $951 ~g$ respectively. 

The current work of estimating the prompt background noise covers $\sim 40\%$ 
of the measured background. The rest part of the background noise is probably
due to the long-term activation of the detector materials due to the CR 
or trapped charged particles which is to be estimated in subsequent work
(Zoglauer et al. 2008; Zoglauer, 2009)

As we have already mentioned that this experiment is primarily designed for
the spectroscopic measurement of Solar flares, but it is also capable of 
detecting GRBs. Since the energy threshold for these GRB detection is about 
$100 ~keV$ (for off-axis source positions) (see, Figure 6b), they will be 
sensitive for GRBs $> 10^{-5} ergs/cm^2$. The probability of on-axis detection 
is less than 1 in 1000 and hence GRBs are not expected  to be detected on-axis.   
As for the uniqueness of the experiment, if we have a reasonable spectral 
response above $100 ~keV$, it is possible to constrain the spectral parameters 
using the data in conjunction with other contemporaneous spectral measurements. 
The exercise of the present paper is to demonstrate that off-axis response can 
be handled in a reasonable way.

On 30th January, 2009, the CORONAS-PHOTON satellite was launched successfully
and all the RT-2 instruments are functioning to our satisfaction. Already
several gamma ray bursts (Rao et al. 2009; Chakrabarti et al. 2009abc) and
solar flares have been detected by the instrument. Detailed reports on the
on-board data quality and backgrounds would be discussed elsewhere. Detailed
results on the observed GRBs and solar flares are also being submitted for
publication elsewhere.


\begin{acknowledgements}
RS and TBK thank RT-2/SRF fellowship (ISRO) which supported their research work.
The authors are thankful to ICSP/TIFR/VSSC/ISRO-HQ for various supports during
RT2 related experiments. We are thankful to an anonymous referee for his very helpful
comments which improved the paper substantially.
\end{acknowledgements}


{}

\end{document}